\documentclass[10pt, conference, compsocconf]{IEEEtran}
\usepackage{cite}
\usepackage{amsmath,amssymb,amsfonts}
\usepackage{algorithmic}
\usepackage{graphicx}
\usepackage{textcomp}
\usepackage[hyphens]{url}

\def\BibTeX{{\rm B\kern-.05em{\sc i\kern-.025em b}\kern-.08em
    T\kern-.1667em\lower.7ex\hbox{E}\kern-.125emX}}

\usepackage{subfig}
\usepackage{braket}
\usepackage{mathtools}
\usepackage{makecell}

\usepackage{physics}
\usepackage{verbatim}

\usepackage[table,xcdraw]{xcolor}

\usepackage{algorithmic}
\usepackage{graphicx}
\usepackage{multirow}
\usepackage{textcomp}
\usepackage{setspace}
\usepackage{authblk}
\usepackage[symbol]{footmisc}

\usepackage[bookmarks=true,breaklinks=true,letterpaper=true,colorlinks,linkcolor=black,citecolor=blue,urlcolor=black]{hyperref}
\usepackage[capitalise]{cleveref} 

\definecolor{revc}{RGB}{0, 0, 0}

\newcommand{\revc}[1]{\textcolor{revc}{#1}}
\newcommand{\todo}[1]{\textcolor{red}{!!!}}

\newcommand{\ubs}{U_{bs}}

\newcommand{\swap}{\text{SWAP}}

\newcommand{\cz}{\text{CZ}}

\newcommand{\rz}{Rz}
\newcommand{\ry}{Ry}

\def\dqbase/{\emph{DigiQ\_baseline}}
\def\dqopt/{\emph{DigiQ\_opt}}
\def\dqmin/{\emph{DigiQ\_min}}
\def\sfqnaive/{\emph{SFQ\_MIMD\_naive}}
\def\sfqmimd/{\emph{SFQ\_MIMD\_decomp}}
\def\sfqsimd/{\emph{SFQ\_SIMD\_decomp}}
\def\dq/{\emph{DigiQ}}

\newcommand{\yy}{\ry(\frac{\pi}{2})}
\newcommand{\uqq}{U_{qq}}

\crefname{section}{Sec.}{Secs.}

\makeatletter
\def\ps@IEEEtitlepagestyle{%
  \def\@oddfoot{\mycopyrightnotice}%
}
\def\mycopyrightnotice{%
  \begin{minipage}{\textwidth}
  \centering \scriptsize
  Copyright~\copyright~2022 IEEE. Personal use of this material is permitted. Permission from IEEE must be obtained for all other uses, in any current or future media, including reprinting/republishing this material for advertising or promotional purposes, creating new collective works, for resale or redistribution to servers or lists, or reuse of any copyrighted component of this work in other works.
  \end{minipage}
}
\makeatother

\begin{document}
\bstctlcite{ieeetr:bstcontrol}

\title{DigiQ: A Scalable Digital Controller for Quantum Computers Using SFQ Logic} 

\author[1]{\rm Mohammad Reza Jokar}
\author[1,3]{\rm Richard Rines}
\author[2,4]{\rm Ghasem Pasandi}
\author[2]{\rm Haolin Cong}
\author[1†]{\rm Adam Holmes}
\author[5]{\rm \\ Yunong Shi}
\author[2]{\rm Massoud Pedram}
\author[1,3*]{\rm Frederic T. Chong}
\affil[1]{Department of Computer Science, University of Chicago}
\affil[2]{Department of Electrical and Computer Engineering, University of Southern California (USC)}
\affil[3]{Super.tech, $^4$NVIDIA, $^5$Amazon Braket
\protect\\ \protect\\ Email: jokar@uchicago.edu, richrines@alum.mit.edu, \{pasandi,haolinco\}@usc.edu, \protect\\ adholmes@uchicago.edu, shiyunon@amazon.com, pedram@usc.edu, chong@cs.uchicago.edu} 

\maketitle
\pagestyle{plain}

\begin{abstract}
\footnote[0]{†Adam Holmes is now at HRL Laboratories, LLC.}\footnote[0]{*Disclosure:  Fred Chong is Chief Scientist at Super.tech and an advisor to Quantum Circuits, Inc.}The control of cryogenic qubits in today's superconducting quantum computer prototypes presents significant scalability challenges due to the massive costs of generating/routing the analog control signals that need to be sent 
from a classical controller at room temperature to the quantum chip inside the dilution refrigerator.
Thus, researchers in industry and academia have focused on designing in-fridge classical controllers in order to mitigate these challenges. Due to the maturity of CMOS logic, many industrial efforts (Microsoft, Intel) have focused on Cryo-CMOS as a near-term solution to design in-fridge classical controllers. 
Meanwhile, Superconducting Single Flux Quantum (SFQ) is an alternative, less mature classical logic family proposed for large-scale in-fridge controllers. SFQ logic has the potential to maximize scalability thanks to its ultra-high speed and very low power consumption. However, architecture design for SFQ logic poses challenges due to its unconventional pulse-driven nature and lack of dense memory and logic. Thus, research at the architecture level is essential to guide architects to design SFQ-based classical controllers for large-scale quantum machines.

In this paper, we present \dq/, the first system-level design of a Noisy Intermediate Scale Quantum (NISQ)-friendly SFQ-based classical controller. We perform a design space exploration of SFQ-based controllers and co-design the quantum gate decompositions and SFQ-based implementation
of those decompositions to find an optimal SFQ-friendly design point that trades area and power for latency and control while ensuring good quantum algorithmic performance.
Our co-design results in a single instruction, multiple data (SIMD) controller architecture, which has high scalability, 
but imposes new challenges on the calibration of control pulses.
We present software-level solutions to address these challenges, which if unaddressed would degrade quantum circuit fidelity given the imperfections of qubit hardware.

To validate and characterize \dq/, we first implement it using hardware description languages and synthesize it using state-of-the-art/validated SFQ synthesis tools.
Our synthesis results show that \dq/ can operate within the tight power and area budget of dilution refrigerators at $>$42,000-qubit scales.
Second, we confirm the effectiveness of \dq/ in running quantum algorithms by modeling the execution time and fidelity of a variety of NISQ applications. We hope that the promising results of this paper motivate experimentalists to further explore SFQ-based quantum controllers to realize large-scale quantum machines with maximized scalability.

\end{abstract}

\begin{IEEEkeywords}
SFQ-based quantum gate; Quantum optimal control; Scalable quantum computer; Cryogenic electronic;
\end{IEEEkeywords}

\IEEEpeerreviewmaketitle

\section{Introduction}
\label{sec:intro}

\begin{figure}[t]
  \begin{center}
  \includegraphics[width=0.9\linewidth]{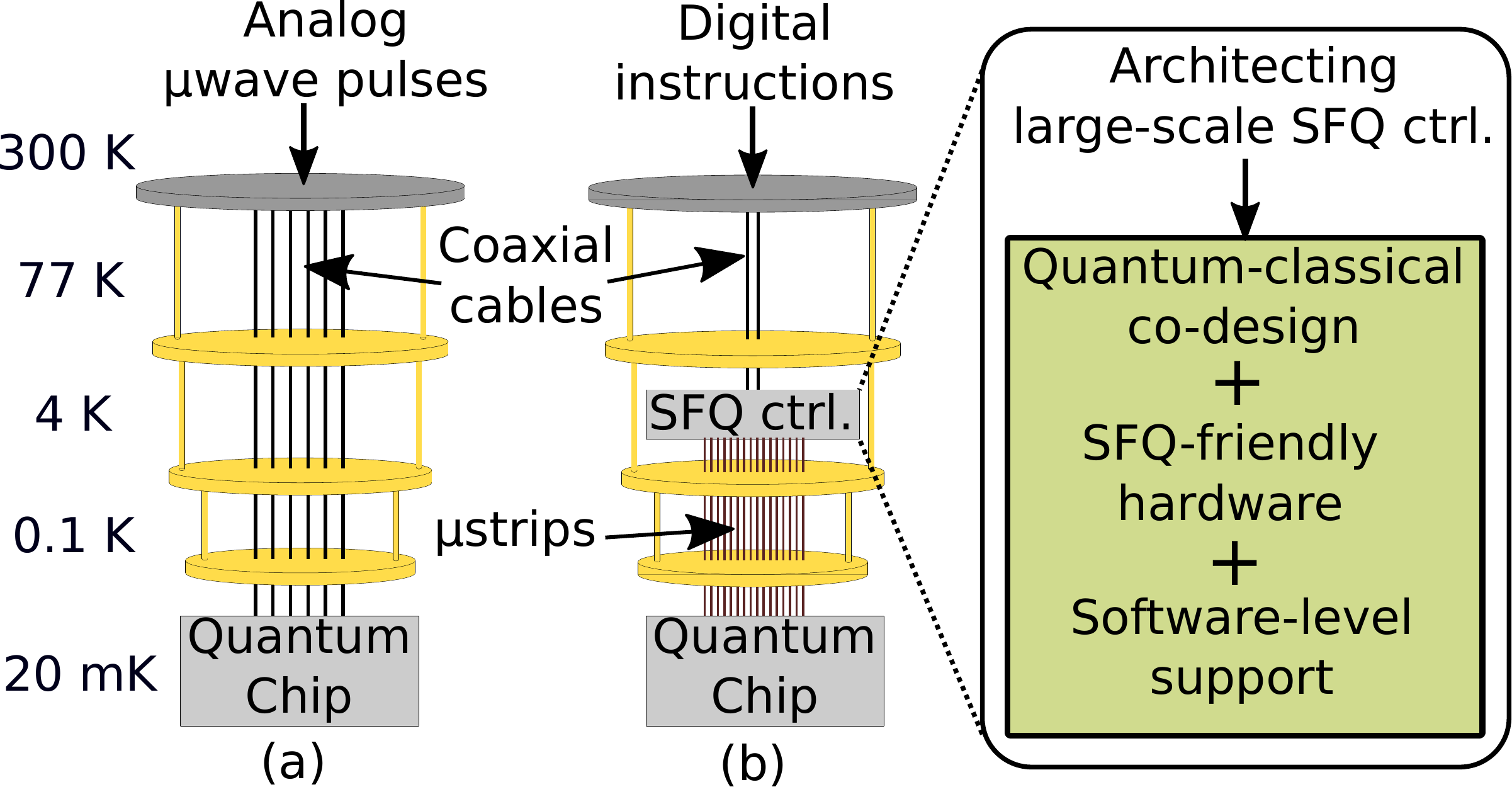}
  \caption{(a) Today's controller design: controller at room temperature, (b) \dq/: controller close to quantum chip.}
  \label{fig:fridge}
  \end{center}
  \vspace{-22pt}
\end{figure}

Superconducting quantum computing \revc{is one of the} promising technologies for building a quantum computer \cite{ibm_device,qarch_yuan}, with many prototypes having been manufactured in the recent years \cite{google_machine1,google_machine2,ibm_machine,qarch_fu}. However, today's prototypes rely on sending separate analog microwave control pulses for each qubit \revc{through coaxial cables} from a classical controller at room temperature to the quantum chip inside the dilution refrigerator (see Fig. \ref{fig:fridge}(a)). This design is simple and straightforward, however presents severe scalability challenges due to the massive costs of generating/routing the analog microwave signals, and significant heat dissipation at millikelvin temperatures due to using a large number of high bandwidth coaxial cables \cite{mcdermott_full,mcdermott_fab,mcdermott_kangbo}.
Thus, it is essential to build compact controllers as close as possible to quantum chips in order to generate and route the control signals locally and address the scalability problem of today's systems (see Fig. \ref{fig:fridge}(b) for an example of such a controller).

Cryo-CMOS is an excellent near-term solution to increase the scalability of today's quantum machines, and controller prototypes based on Cryo-CMOS have been manufactured \cite{horseridge2} which can scale to $>$800 qubits (see Sec. \ref{sec:related}). Meanwhile, Superconducting Single Flux Quantum (SFQ) logic \cite{sfq_logic1,sfq_logic2} is an emerging classical logic family and is recognized as a promising solution to maximize the scalability of in-fridge controllers due to its unique characteristics such as ultra-high speed and very low power \cite{mcdermott_fab,mcdermott_full,mcdermott_kangbo,sfq_genetic}. However, a key remaining step is designing an SFQ-friendly controller architecture that operates within the tight power and area budget of dilution refrigerators at large scales, while ensuring good quantum algorithmic performance.

Prior work has demonstrated quantum gates based on SFQ logic \cite{mcdermott_fab,mcdermott_kangbo,sfq_genetic},
and has outlined a vision for an SFQ-based controller for fault-tolerant quantum computing that relies on repeated streaming of quantum instructions that are stored in SFQ registers \cite{mcdermott_full}.
These studies done by physicists are essential in order to show the feasibility of performing SFQ-based quantum gates, and envision the potentials of SFQ logic in controlling large-scale quantum chips. However, the following architectural shortcomings remain to be addressed: (1) prior work does not consider scenarios where we are no longer primarily repeating the same quantum instructions. Thus, they are especially not suitable for running Noisy Intermediate Scale Quantum (NISQ) algorithms;
(2) there has not been a detailed synthesis of a complete SFQ-based controller architecture in order to get an accurate estimate of power and area, and determine the scale we can achieve with such controllers given the power and area budget of dilution refrigerators; 
(3) there has not been an analysis of the implications of SFQ-based quantum controllers on quantum algorithmic performance.
This analysis is necessary to assess the cost and effectiveness of SFQ-based controller designs; (4) there has not been an analysis on the robustness of SFQ-based controllers to the qubit imperfections in NISQ systems.

In this paper, we address the above architecture-level shortcomings and present \dq/, the first system-level design of a scalable NISQ-friendly SFQ-based classical controller.
Inspired by the SuperNPU paper \cite{supernpu}, which demonstrates architecture design for SFQ-based neural processing units, our paper guides architects to design SFQ-based controller architectures for large-scale quantum computers. 

Architecture design for SFQ logic is different from that of CMOS logic, due to its unconventional pulse-driven nature and lack of dense memory/logic. In addition, implementation of quantum gates using SFQ pulses is different from that of microwave pulses. Thus, novel SFQ-friendly controller architecture designs are essential. We perform a design space exploration of SFQ-based controllers and co-design the quantum gate decompositions and SFQ-pulse implementation of those decompositions
to ensure that our design both works within the tight power and area budget of dilution refrigerators at large scales and provides good algorithmic performance.

\revc{Quantum gate parallelism is essential to preserve good quantum algorithmic performance in many NISQ applications \cite{nisq_parallel}.}
Our quantum-classical co-design demonstrates that due to the lack of dense memory/logic in SFQ and tight power and area budget of dilution refrigerators, \revc{we cannot afford to simultaneously send tailored quantum gates to many qubits at large scales. The implementation of quantum algorithms with significant gate parallelism therefore requires sharing SFQ-based quantum instructions among multiple qubits (i.e., single instruction, multiple data (SIMD)).
Getting good algorithmic performance on a SIMD NISQ architecture is especially challenging because of qubit variations and frequency drift, which typically require gates to be uniquely calibrated for each qubit.
We propose novel software-level solutions to address these challenges and preserve SIMD parallelism.}

To validate and characterize \dq/, we first work out the details of quantum program execution flow, starting from our compiler at room temperature and ending with sending the control signals to qubits. Then, we implement our complete design in hardware description languages, and synthesize it using state-of-the-art/validated SFQ synthesis tools \cite{coldflux,pasandi2019dynamic,pasandi2019efficient,shahsavani2017integrated} to measure power, area, and delay values. Our synthesis results are obtained post-layout based on accurate extraction of the gate layouts and passive transmission lines and subsequently simulated using well-established tools for superconductive electronic applications \cite{coldflux,wrspice,jsim}, and are thus highly accurate and reliable. Finally, we show the effectiveness of \dq/ in running quantum algorithms by compiling a variety of NISQ algorithms for our system, and modeling the resulting execution time and fidelity. 

\revc{We position ourselves as addressing the physical challenges in scaling up the NISQ machines. Our work is complementary to Perfect Intermediate Scale Quantum computing (PISQ) \cite{pisq} approach, which suggests developing new quantum algorithms assuming perfect qubits and evaluating them with classical simulators. PISQ is a great approach since it separates the development of quantum algorithms and applications from hardware and architectural research and allows researchers to focus on the development of new quantum algorithms in various scientific fields.}

To summarize, our key contributions are as follows:

\begin{itemize}
    \item \textbf{Architecting SFQ-based controllers:} We guide architects to design large-scale SFQ-based controllers by taking into consideration the opportunities (efficient on-chip broadcast and ultra-fast clock) and challenges (lack of dense memory/logic) of SFQ.
    \item \textbf{Quantum-classical co-design:} By co-designing 
    quantum gate decompositions and SFQ-based implementation of those decompositions, we present \dq/, an SFQ-friendly SIMD controller architecture.
    \item \textbf{Addressing the SIMD calibration challenges:} We present software solutions to address the quantum gate calibration challenges of SIMD hardware to make it robust to imperfections in qubit hardware.
    \item \textbf{Characterizing controller hardware:} We implement \dq/ in hardware description languages and show its feasibility in large scales in terms of power, area, and delay using state-of-the-art/validated SFQ synthesis tools.
    \item \textbf{Characterizing algorithmic performance:} We confirm the effectiveness of \dq/ in running quantum algorithms by compiling a variety of NISQ applications for \dq/ and modeling their execution time and fidelity.
\end{itemize}

The rest of the paper is organized as follows. Sec. \ref{sec:back} provides a background on quantum controllers, SFQ logic, and discusses the opportunities and challenges of SFQ-based controllers. We present related work in Sec. \ref{sec:related}. Sec. \ref{sec:scheme} demonstrates the details of \dq/, our scalable digital SFQ-based quantum controller architecture. Sec. \ref{sec:simd} discusses the quantum gate calibration challenges of \dq/, and presents software-level solutions to address them. Sec. \ref{sec:results} shows our methodology and thorough evaluation of \dq/. Finally, Sec. \ref{sec:conclusion} concludes the paper and discusses the future work.

\section{Background and Motivation}
\label{sec:back}

Here we provide a background on quantum computing followed by a discussion of quantum gates/controllers in existing systems and their limitations. We then present a background on SFQ logic and discuss the opportunities and challenges of SFQ-based controllers. 

\subsection{Quantum computing}
\begin{figure}[t!]
    \centering
    \includegraphics[width=0.4\textwidth]{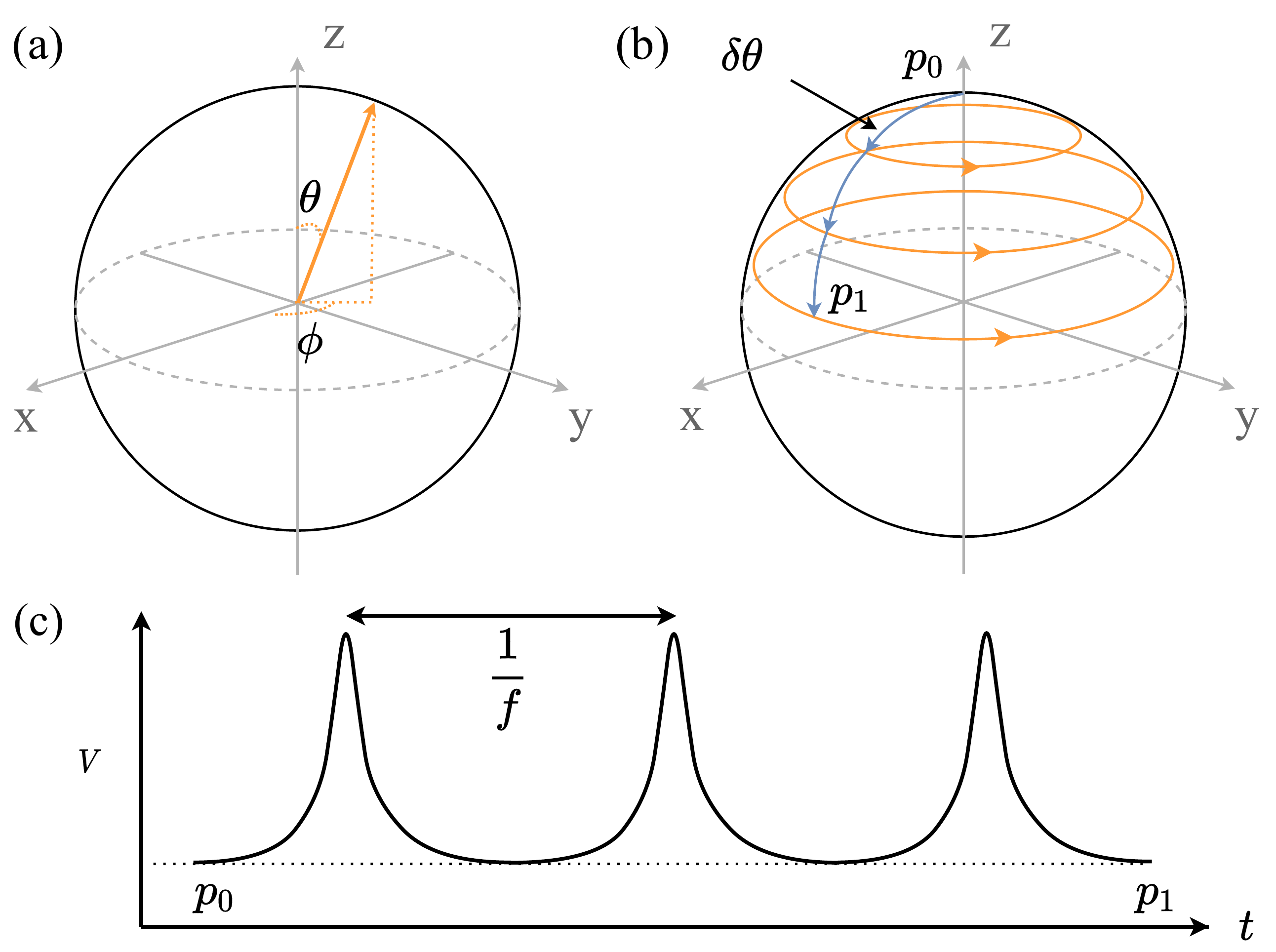}
    \caption{(a) Bloch sphere representation of a qubit; 
    (b) SFQ driven trajectory. The blue trajectory is driven by the periodic SFQ pulse train shown in (c), and the orange trajectory is driven by the qubit free evolution; (c) SFQ pulse train in the time domain.
    $f$ is the qubit oscillation frequency. 
    }
    \label{fig:sfq_single_q}
    \vspace{-12pt}
\end{figure}

A quantum algorithm specifies a series of transformations on quantum systems called qubits, which are analogous information carriers to classical bits. The state of a single qubit can be represented as a linear combination of two states:
\begin{align*}
    \ket{\psi} = \alpha \ket{0} + \beta \ket{1}
\end{align*}
where the {amplitudes} $\alpha, \beta \in \mathbb{C}$ satisfy $|\alpha|^2 + |\beta|^2 = 1$.
It is useful to visualize the state of a single qubit as a vector on the unit Bloch sphere as shown in Fig. \ref{fig:sfq_single_q}(a), where $\alpha=\cos\theta/2$ and $\beta=e^{i\phi}\sin\theta/2$.
A multi-qubit state may be written as
    $\ket{\psi} = \sum_i \alpha_i \ket{i}$,
where $\sum_i|\alpha_i|^2=1$ and $\ket{i}=\ket{i_{n-1}}\ldots\ket{i_1}\ket{i_0}$ are the computational basis states of the $n$-qubit quantum system. 

\emph{Quantum gates} are unitary operators which modify the state of the qubit system. Any single-qubit gate can be represented as a rotation on the unit Bloch sphere (see Fig. \ref{fig:sfq_single_q}(a)). Rotations around any two axes can be combined to perform arbitrary single-qubit gates. Combined with
a two-qubit \emph{entangling} gate (i.e., a gate which cannot be decomposed into one-qubit gates), this is sufficient for universal quantum computation \cite{Barenco1995}.

\subsection{Superconducting qubit controllers and their limitations}
\label{sec:background:control}

Superconducting qubits are nonlinear LC circuits built with Josephson junctions (JJ) that operate at near absolute zero temperature and oscillate at microwave frequencies.
Qubits are defined using the quantized ground ($\ket0$) and first excited ($\ket1$) states of the oscillator, where the qubit's oscillation frequency is defined as the energy difference between the two levels.
\emph{Transmons} are a simple form of superconducting qubit comprising a JJ and a shunt capacitor, designed to reduce the qubit's sensitivity to electrical charge noise~\cite{transmon}, and is the qubit technology in many state-of-the-art systems \cite{google_machine1}.
\emph{Flux-tunable} transmons allow tuning the qubit oscillation frequency, and are implemented by replacing the transmon's JJ with a pair of parallel junctions separated by a small distance. The qubit frequency can then be shifted by driving an external magnetic flux (using a small electrical current, $\sim$1 mA) through the enclosed loop \cite{quantum_eng}. 
The relationship between frequency and flux can be fine tuned by varying the parameters of each JJ individually, creating what's known as an \emph{asymmetric} transmon \cite{hutchings2017}.

Superconducting quantum computer prototypes perform single-qubit gates by driving transitions between the oscillator's $\ket0$ and $\ket1$ states using microwave pulses.
Controllable multi-qubit operations require a means of selectively interacting qubits through some coupling architecture.
Two-qubit gates such as
$\cz$ can be implemented using flux-tunable transmons by changing the frequency of the qubits temporarily such that their quantum states are on resonance \cite{quantum_eng}. 

\revc{Superconducting quantum computers have received significant attention due to their convenient qubit design and configurability \cite{quantum_eng}, leading to rapid advances in their size, coherence time, and gate fidelity \cite{google_machine1,google_machine2,ibm_machine}.
However, they also pose significant challenges which must be addressed for superconducting quantum computing to be scalable.
Superconducting qubits need cooling to very low temperatures ($\sim$20 mK) which is expensive and complicates control hardware. Further, unlike technologies such as trapped ion (in which qubit uniformity is guaranteed by nature \cite{trappedidentical}), the characteristics of superconducting qubits are subject to variation and drifts over time.} 

The controller design in today's prototypes relies on separate cables from room temperature to control individual qubits \cite{google_machine1}. This approach has severe scalability issues. First, there is a massive electronics cost for generating and routing the analog control pulses from room temperature including the cost of arbitrary waveform generators (AWGs), microwave sources, IQ mixers (which modulate the in-phase and out-of-phase components of the drive signal \cite{quantum_eng}),
and amplifiers and attenuators that are used for thermal property matching at each stage of the fridge \cite{mcdermott_fab, quantum_eng, mcdermott_full}. Second, the cooling power at millikelvin stage of the fridge is very limited ($<$10 $\mu$W \cite{qarch_hornibrook}) and a large number of high bandwidth coaxial cables create a big heat load problem at this stage. Thus, alternative quantum controller architectures are needed to realize scalable and robust quantum machines.

\subsection{Opportunities and challenges of SFQ quantum controllers}
Superconducting Single Flux Quantum (SFQ) is a magnetic-pulse based device and a single quantum of magnetic flux, $\Phi_0$ = $h/2e$ = $2.07$ mV$\cdot$ps, is used for logic bit representation. In this representation, presence of a pulse has the meaning of ``logic-1", while absence of a pulse is considered as a ``logic-0". Operation of SFQ logic is based on overdamped JJs \cite{herr2011ultra,takeuchi2013adiabatic,sfq_logic2,sfq_logic1,volkmann2013experimental}.
There are two types of D-Flip-Flops (DFFs) in this technology: Destructive Read Out (DRO) DFF and Non-Destructive Read Out (NDRO) DFFs. In the first type the stored data will be erased after one read operation and the DFF will be reset back to its zero state. However, in the second type, multiple read operations can be done without erasing the content of DFF. 
SFQ devices with switching delay of {1 ps} and switching energy of $10^{-19}$ J are considered great candidates to provide high speed solutions post-silicon and post-CMOS \cite{supernpu}.
Moreover, these Niobium-based devices are extremely low power and despite their cryo-cooling overhead they still consume significantly less power compared to the state-of-the-art silicon-based devices \cite{mukhanov2011energy}. These unique properties make SFQ an attractive logic technology to implement classical controllers with maximized scalability for quantum computers. 

SFQ can be utilized not only to implement classical controller logic, but also to perform quantum gates locally. Prior work demonstrated that SFQ pulse trains can be utilized to perform single-qubit gates \cite{mcdermott2014,sfq_genetic, mcdermott_fab}.
For example, an intuitive approach to do rotations along the y-axis, $Ry(\theta)$, is to apply one SFQ pulse every qubit oscillation period as shown in Fig. \ref{fig:sfq_single_q}. The time integral of an SFQ pulse is equal to the superconducting flux quantum and determines the energy deposited in the qubit. The result of this energy deposition is a small rotation of $\delta \theta$ around the y-axis. We can perform a rotation along the y-axis by keep applying one SFQ pulse every qubit period. An arbitrary single-qubit gate can be represented by a bitstream (denoted as SFQ bitstream); the gate is applied by applying the SFQ bitstream to the qubit, one bit at a time: if the bit is 1 (0), we apply (don’t apply) an SFQ pulse to the qubit at the corresponding SFQ chip cycle. 

Despite its high potentials, SFQ imposes unique constraints on the controller design. First, limited JJ density in existing technology (100X-10,000X \revc{lower density} than CMOS \cite{qarch_swamit2}) leads to lack of dense memory/logic in SFQ. Second, SFQ design is different from that of CMOS due to unconventional pulse-driven nature of SFQ logic. 
SFQ logic gates receive clock signals and they are evaluated by arrival and consumption of clock pulses. Therefore, balancing the circuit path by inserting extra DFFs is essential to ensure that inputs are consumed at correct clock cycles \cite{pasandi2019dynamic}.
The extra DFFs further increase the logic area (and power), and might incur significant costs in complex logic designs. Thus, we need to take these constraints into consideration and (1) use limited SFQ storage; (2) keep the logic simple.

\section{Related Work}
\label{sec:related}
\begin{table*}[ht]
\caption{
\revc{Design space for SFQ-based single-qubit gate controllers.}}
\label{tab:mimdsimd}
\centering
\small
\begin{tabular}{|l|c|c|c|c|}
\hline
 &  &  & \multicolumn{2}{c|}{SFQ\_SIMD\_decomp (DigiQ)} \\ \cline{4-5} 
\multirow{-2}{*}{} & \multirow{-2}{*}{SFQ\_MIMD\_naive} & \multirow{-2}{*}{SFQ\_MIMD\_decomp} & DigiQ\_min & DigiQ\_opt \\ \hline
Scalability & \cellcolor[HTML]{F55955}\begin{tabular}[c]{@{}c@{}}Limited by \\ power, area, and bandwidth\end{tabular} & \cellcolor[HTML]{F55955}\begin{tabular}[c]{@{}c@{}}Limited by \\ power and area\end{tabular} & \multicolumn{2}{c|}{\cellcolor[HTML]{67FD9A}High scalability} \\ \hline
\begin{tabular}[c]{@{}l@{}}Quantum program\\  execution\end{tabular} & \cellcolor[HTML]{67FD9A}\begin{tabular}[c]{@{}c@{}}No gate serialization\\ \end{tabular} & \cellcolor[HTML]{67FD9A}\begin{tabular}[c]{@{}c@{}}No gate serialization\\ \end{tabular} & \cellcolor[HTML]{FFFE65}\begin{tabular}[c]{@{}c@{}}Long\\ decompositions\end{tabular} & \cellcolor[HTML]{FFFE65}\begin{tabular}[c]{@{}c@{}}Potential\\ serialization\end{tabular} \\ \hline
Pulse calibration & \cellcolor[HTML]{67FD9A}Hardware & \cellcolor[HTML]{67FD9A}Hardware & \multicolumn{2}{c|}{\cellcolor[HTML]{FFFE65}Software} \\ \hline
\end{tabular}
\vspace{-10pt}
\end{table*}

In this section we discuss prior research on in-fridge quantum gates and controllers based on various technologies, and also SFQ-based accelerators.

\subsection{Cryo-CMOS based quantum controllers} Due to maturity of CMOS technology, Cryo-CMOS is an attractive technology to do computation at the 4 K stage of the fridge. 
In \cite{horseridge2}, single-qubit gate operation using a Cryo-CMOS controller prototype is demonstrated, which is done by generating microwave control signals inside the fridge using the pulse information that is stored in on-chip SRAMs. The prototype presented in \cite{horseridge2}
can scale to $>$800 qubits given 12 mW/qubit power consumption reported in the paper and 10 W power budget \cite{mcdermott_full}. In comparison, SFQ logic can potentially maximize the scalability of quantum controllers ($>$42,000 qubits in our SIMD design) due to its very low power consumption. Note that 
SIMD can potentially increase the scalability of today's Cryo-CMOS prototypes as well, which we see as important future work (see Sec. \ref{sec:conclusion}).
  
\subsection{SFQ-based quantum gates and controllers} 
In \cite{mcdermott_fab}, coherent control of a qubit using SFQ pulse trains is demonstrated using a DC/SFQ converter that is fabricated on the qubit chip; the experimental results in the paper show the feasibility of performing quantum gates using SFQ. In \cite{mcdermott_kangbo}, the authors propose a method to find SFQ bitstreams for qubits with different frequencies using one single SFQ clock. A genetic algorithm is used in \cite{sfq_genetic} as an approach to find efficient SFQ
bitstreams to reduce leakage errors (i.e., unwanted energy transfer to states other than $\ket{0}$ and $\ket{1}$) and gate time. Reinforcement learning is another approach that has been utilized to find efficient SFQ bitstreams to perform quantum gates \cite{alphazero}. In \cite{mcdermott_full}, the authors outline a vision to perform fault tolerant quantum computing that relies on repeated streaming of stored SFQ bitstreams, and present a simple estimation of power by adding the power of SFQ registers. In contrast to these works, \dq/ is the first system-level design of a scalable NISQ-friendly SFQ-based controller.

\subsection{SFQ-based accelerators}
SFQ has been utilized to design hardware accelerators thanks to its ultra-high speed. In \cite{nisq+}, the authors propose an SFQ-based error decoder to accelerate quantum error correction. In \cite{supernpu}, the authors present an ultra-fast SFQ-based neural processing unit. In contrast to these works, our focus is on designing a scalable SFQ-based controller rather than accelerating a task. 
\section{DigiQ quantum controller}
\label{sec:scheme}

In this section we provide guidelines for designing scalable SFQ-based controller architectures for universal quantum computation, and present \dq/, our novel SFQ-based controller architecture.

\subsection{SFQ-based universal quantum computation}
 
\subsubsection{Design space for single-qubit gate controllers}
\label{sec:scheme:one-qubit}
 
A naive design to do arbitrary single-qubit gates is to allocate separate SFQ registers for SFQ bitstreams per qubit (similar to \cite{mcdermott_full}) and update them as needed. This design, denoted as \sfqnaive/, is similar in spirit to today's microwave-based designs, but relies on digital communication from room temperature rather than analog communication. \sfqnaive/ is straightforward and provides quantum gate parallelism (i.e., multiple instruction, multiple
data (MIMD) paradigm). However, it requires high communication bandwidth from room temperature in scenarios where we are not primarily repeating the same quantum gates, and thus need to update a large number of SFQ registers on-the-fly during the quantum program execution.
In addition, \sfqnaive/ is expensive in terms of power and area (5.01 mW/qubit and 13.9 mm$^2$/qubit just for 300-bit SFQ registers based on our results obtained using detailed SFQ synthesis tools and cell libraries).
Thus, the scalability of \sfqnaive/, is limited by power/area/bandwidth.

In order to reduce the required bandwidth from room temperature, we can store a universal single-qubit gate set per qubit on the SFQ chip, and send the quantum gate decomposition information from the room temperature. In the simplest case where the single-qubit gate set includes two gates, we need to send only 1 bit per qubit at any given time from room temperature to select/apply one of the two gates. This design, denoted as \sfqmimd/, reduces the bandwidth requirement significantly compared to \sfqnaive/. However, it further increases the power and area as it allocates more than one SFQ register per qubit. Thus, the scalability of \sfqmimd/ is still limited by power/area. 

Finally, we demonstrate our scalable design. Tight power, area, and bandwidth budget of dilution refrigerators, and lack of dense memory in SFQ (see Sec. \ref{sec:back}) leads us to a design where we share SFQ bitstreams among a group of qubits (i.e., SIMD). \revc{Grouping is static and done at the design time, such that qubits in a group have the same nominal oscillation frequency; we show that we can compensate for frequency drift in software in \cref{sec:simd}.} Note that sharing the bitstreams can be done efficiently in SFQ by broadcasting the bitstreams (one bit per SFQ cycle) using splitter gates. This design, denoted as \sfqsimd/, is a potential candidate to realize a controller with high scalability thanks to its reduced power, area, and bandwidth requirements compared to the other two designs. We therefore base \dq/ on this design.
Table \ref{tab:mimdsimd} summarizes the investigated design space. 

\subsubsection{Single-qubit gate decomposition}
Out of the plethora of proposed quantum gate decomposition protocols for single-qubit gates, we must choose the ones that can be implemented efficiently using SFQ. We prefer a decomposition protocol that relies on storing/processing a limited number of SFQ bitstreams (see Sec. \ref{sec:back}). This is important because (1) there is a lack of dense memory in SFQ, thus we can afford to store only a limited number of SFQ bitstreams; (2) we need to keep the SFQ logic simple, thus we can afford to process only a limited number of SFQ bitstreams at any given time. We consider two SFQ-friendly gate decompositions and present their corresponding SFQ-based architecture designs.

Our first architecture, denoted as \dqmin/, is based on a minimal universal single-qubit gate set that includes only 2-4 gates (e.g., $[Ry(\frac{\pi}{2}),T]$ is sufficient for universal single-qubit computation \cite{kitaev1997,solovaykitaev}) with the goal of minimizing the hardware cost. Our digital controller works with a clock (denoted as \emph{quantum controller clock}), and similar to classical computers, each quantum program is decomposed into a number of quantum gates and each quantum gate is finished in a number of controller cycles. In \dqmin/, single-qubit gates are decomposed into a sequence of the gates in the gate set, and each gate of the sequence is executed in one controller cycle. At the beginning of each controller cycle, the SFQ bitstreams corresponding to all the gate are broadcasted to a group of qubit controllers. Each qubit controller uses a simple SFQ-based multiplexer to select/apply one of the possible bitstreams using the control bits that are sent from the room temperature in each controller cycle.

A concern is that \dqmin/ needs long gate sequences to perform arbitrary single-qubit gates.
Thus, we also present \dqopt/, in which 
we implement the continuous gate set $\{Ry(\frac{\pi}{2}),Rz(\phi);\phi\in[0,2\pi)\}$, enabling constant-depth single-qubit gate decomposition \cite{virtualz}. 
$Ry(\frac{\pi}{2})$ can be implemented by storing an SFQ bitstream on the SFQ chip.
The next question is: how to implement $Rz(\phi)$ gates for arbitrary $\phi$ efficiently?

In microwave-based systems, $Rz(\phi)$ gates can be done virtually (in software) by changing the phase of the microwave pulses of the consecutive gates \cite{quantum_eng,virtualz}. But, in the SFQ case, we do not have control over the phase of the SFQ bitstream. Thus, we need to perform the $Rz(\phi)$ gates in hardware. Thanks to the ultra-fast, precise clock on the SFQ chip (in the order of ps), we can perform arbitrary $Rz(\phi)$ by letting the qubits evolve freely for a precise number of SFQ chip clock cycles, which is equivalent to applying an SFQ bitstream consisting of only ``0'' bits. 
Every qubit oscillation period, a qubit performs a full rotation around the z-axis with a trajectory in the form of a unit circle (see Fig. \ref{fig:sfq_single_q}). Now, for any $\phi$, there is a point on the unit circle corresponding to $Rz(\phi)$ and we need to get as close as possible to that point in order to perform $Rz(\phi)$ with high fidelity. 
By applying a ``0'' bit every SFQ chip cycle, we get to multiple points on the unit circle in one qubit oscillation period, and we cover more points if we let the qubit evolve freely for more than one qubit period. 
The longer we let the qubit evolve freely, the more points we can get to on the unit circle and the more precisely we can approximate $Rz(\phi)$ for arbitrary $\phi$.
The fidelity of these gates is analyzed in Sec. \ref{sec:simd:1qb}.

\begin{figure}[t!]
  \begin{center}
  \includegraphics[width=0.98\linewidth]{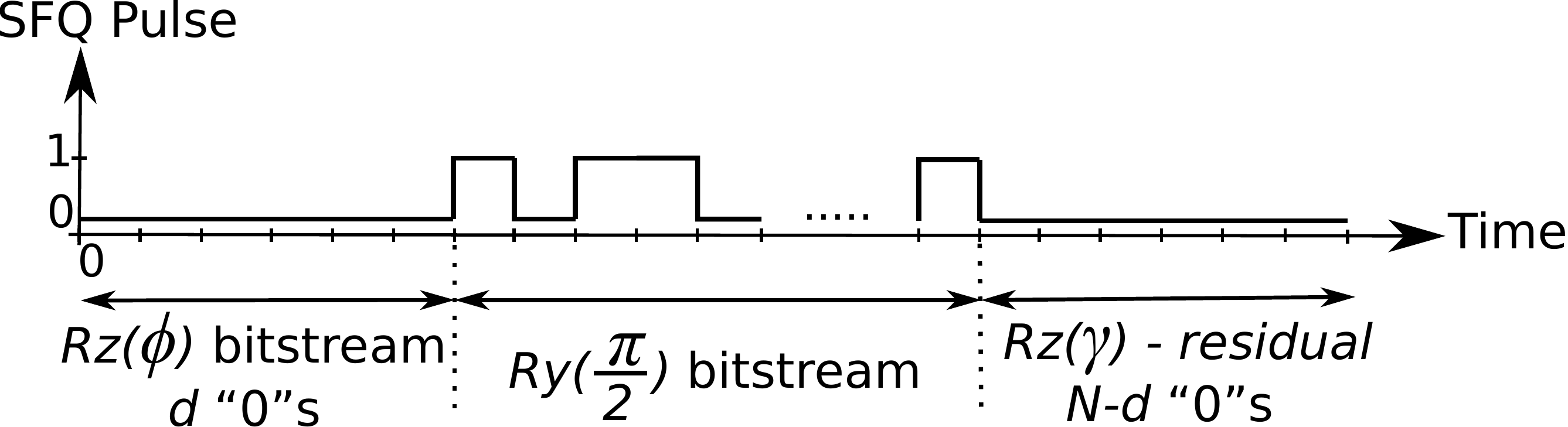}
  \caption{The sequence of gates in one cycle of \dqopt/.}
  \label{fig:digibase}
  \end{center}
  \vspace{-10pt}
\end{figure}

We now discuss implementation details of \dqopt/ single-qubit gates. 
Any one-qubit gate can be decomposed as $U(\phi_3,\phi_2,\phi_1) = Rz(\phi_3)Ry(\frac{\pi}{2})Rz(\phi_2)Ry(\frac{\pi}{2})Rz(\phi_1)$.
In \dqopt/, we break the $U(\phi_3,\phi_2,\phi_1)$ into three parts, $Ry(\frac{\pi}{2})Rz(\phi_1)$, $Ry(\frac{\pi}{2})Rz(\phi_2)$, and $Rz(\phi_3)$. We perform each of the first two parts in one controller cycle, 
and absorb the $Rz(\phi_3)$ to the next controller cycles. The controller cycle length is equal to the $Ry(\frac{\pi}{2})$ gate time plus $N*SFQ\_chip\_cy\-cle\_length$, where $N$ is the maximum number of SFQ chip cycles that we let the qubits evolve freely in order to perform $Rz(\phi)$ gates. We perform $Ry(\frac{\pi}{2})Rz(\phi)$ by applying a ``0'' bitstream of length $0 \leq d \leq N$ followed by the $Ry(\frac{\pi}{2})$ bitstream, which is equivalent to delaying the $Ry(\frac{\pi}{2})$ bitstream by $d$ SFQ chip cycles as shown in Fig. \ref{fig:digibase}. The value of $d$ depends on the angle $\phi$ and is sent by the compiler at room temperature every controller cycle; the SFQ logic in \dqopt/ delays the stored $Ry(\frac{\pi}{2})$ bitstream by $d$ SFQ chip cycles and broadcasts it to a group of qubit controllers. The residual $Rz(\gamma)$ rotation at the end of the controller cycle (see Fig. \ref{fig:digibase}) can be absorbed into the next gate on that qubit.

The next question is: how many distinct $Ry(\frac{\pi}{2})Rz(\phi)$ gates out of a total of $N+1$ possible gates (denoted as $BS$) are needed every controller cycle? The answer depends on (1) the available power and area budget inside the fridge: providing more distinct gates requires more complex logic to delay the stored $Ry(\frac{\pi}{2})$ bitstream by more distinct values at the same time, and also broadcast and process more SFQ bitstreams; (2) the similarity between the gates that qubits inside a group perform in a given instant: there will be serialization inside a group if not enough distinct gates are available in a controller cycle. The lower the $BS$ value, the lower the hardware cost but the higher the chance of serialization (see Sec. \ref{sec:results}).

\begin{figure}[t!]
  \centering
  \begin{tabular}[b]{c}
    \includegraphics[width=.8\linewidth]{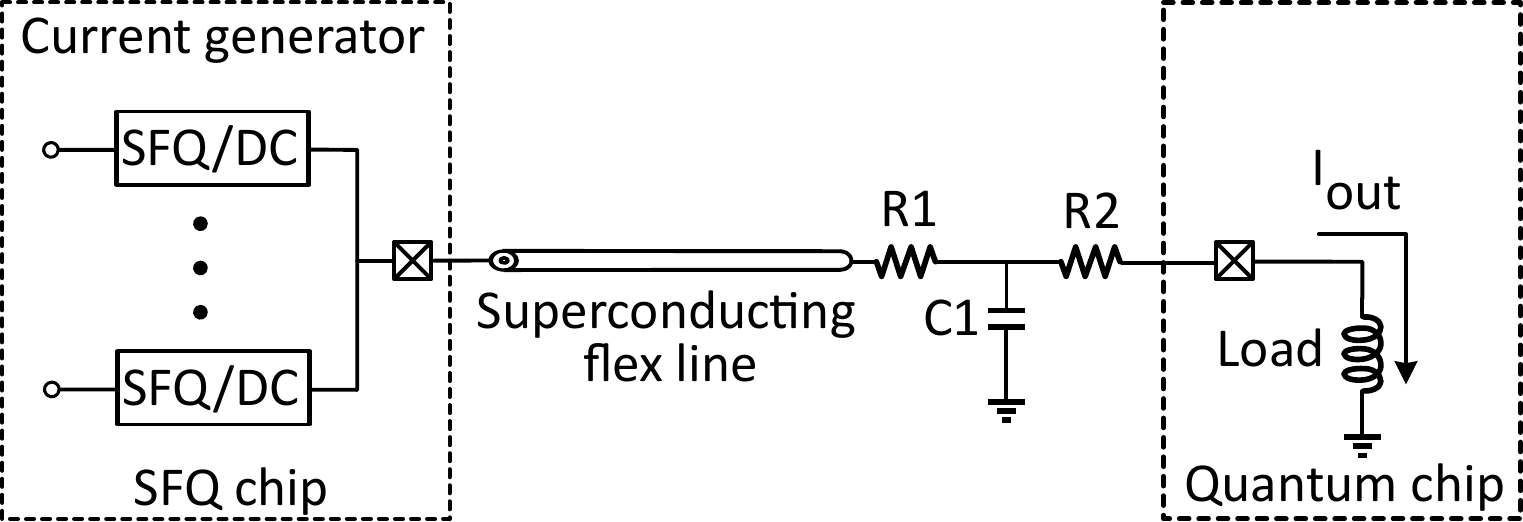} \\
    \small (a)
  \end{tabular} \qquad
  \begin{tabular}[b]{c}
    \includegraphics[width=.7\linewidth]{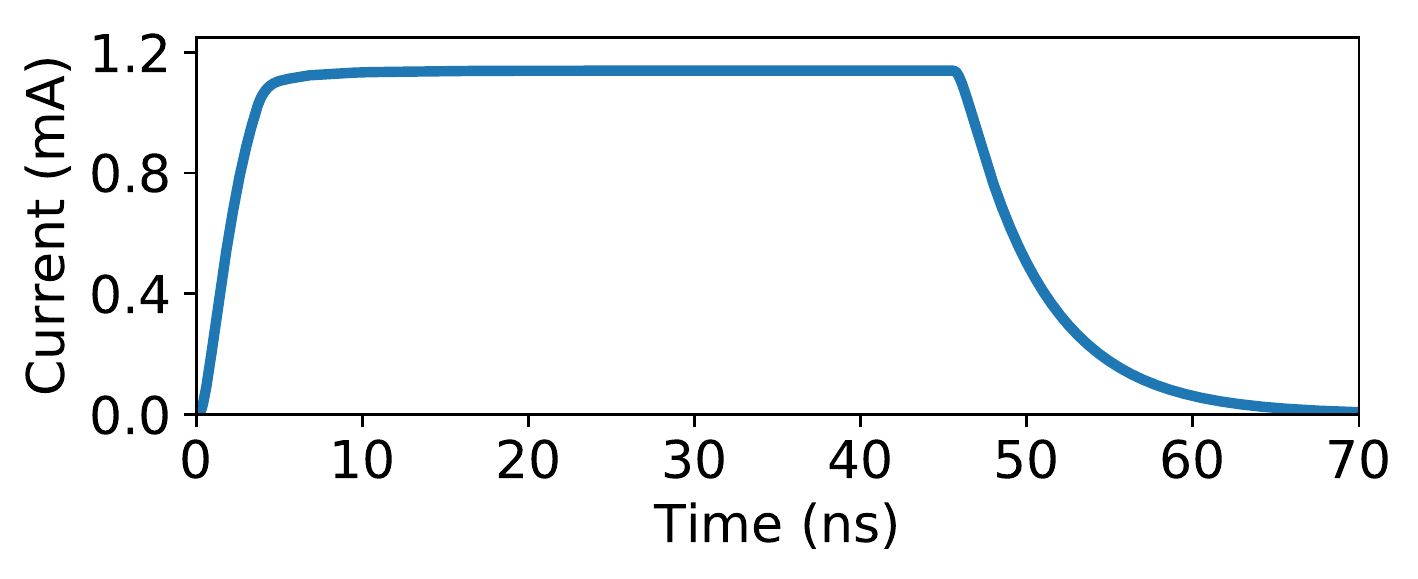} \\
    \small (b)
  \end{tabular} 
  \vspace{-10pt}
  \caption{(a) Circuit schematic of our current generator design based on SFQ/DCs; (b) The electrical current pulse generated by our design to realize $\cz$ gates on flux-tunable transmons.}
  \label{fig:CZ}
  \vspace{-12pt}
\end{figure}

\subsubsection{Two-qubit gate design}
\label{sec:scheme:2qubit}
In this paper, we develop SFQ circuits to generate electrical current waveforms necessary to realize $\cz$ gates on flux-tunable transons (see Sec. \ref{sec:back}) inside the fridge. The electrical currents are transmitted to the quantum chip using superconducting microstrip flex lines (see Fig. \ref{fig:CZ}(a)). 
We deploy an array of SFQ/DCs, which are commonly used SFQ blocks to convert the SFQ pulsed representation to DC voltage levels \cite{sfqdc}.
When a \emph{start} signal is received by the SFQ/DC blocks, they start outputting electrical current, and they keep doing so until they receive a \emph{stop} signal (we need to turn the SFQ/DCs on (off) in the beginning (end) of the $\cz$ gate).
In order to target specific pairs of qubits on a multi-qubit system we require current waveforms to be applied to both qubits simultaneously (we need one current generator per qubit). 
The only difference between this approach and prior flux-tunable implementation of CZ gates is that the electrical current is generated inside the fridge. Note that we use the same two-qubit gate design for both \dqmin/ and \dqopt/.

We use JSIM which is a detailed circuit simulator for superconductive electronic applications \cite{jsim} to simulate the Fig. \ref{fig:CZ}(a) current generator circuit. Fig. \ref{fig:CZ}(b) shows the simulated current waveform,
which is generated by enabling 25 SFQ/DC blocks (the values of $R1$, $R2$, and $C1$ are 0.05 ohm, 0.05 ohm, and 10 nF, respectively).
In Sec. \ref{sec:simd:2qb}, we use physical simulations to show that this waveform can be combined with software calibration techniques to realize low-error CZ gates even when qubits exhibit independent frequency variation.

Recent studies have also suggested realizing two-qubit cross-resonance gates using only SFQ pulses \cite{jokarqce,alphazero}.
The design of an SFQ controller architecture based on such gates would present a different set of challenges and opportunities, the study of which we save for future work.

\begin{figure}[t!]
  \begin{center}
  \includegraphics[width=0.95\linewidth]{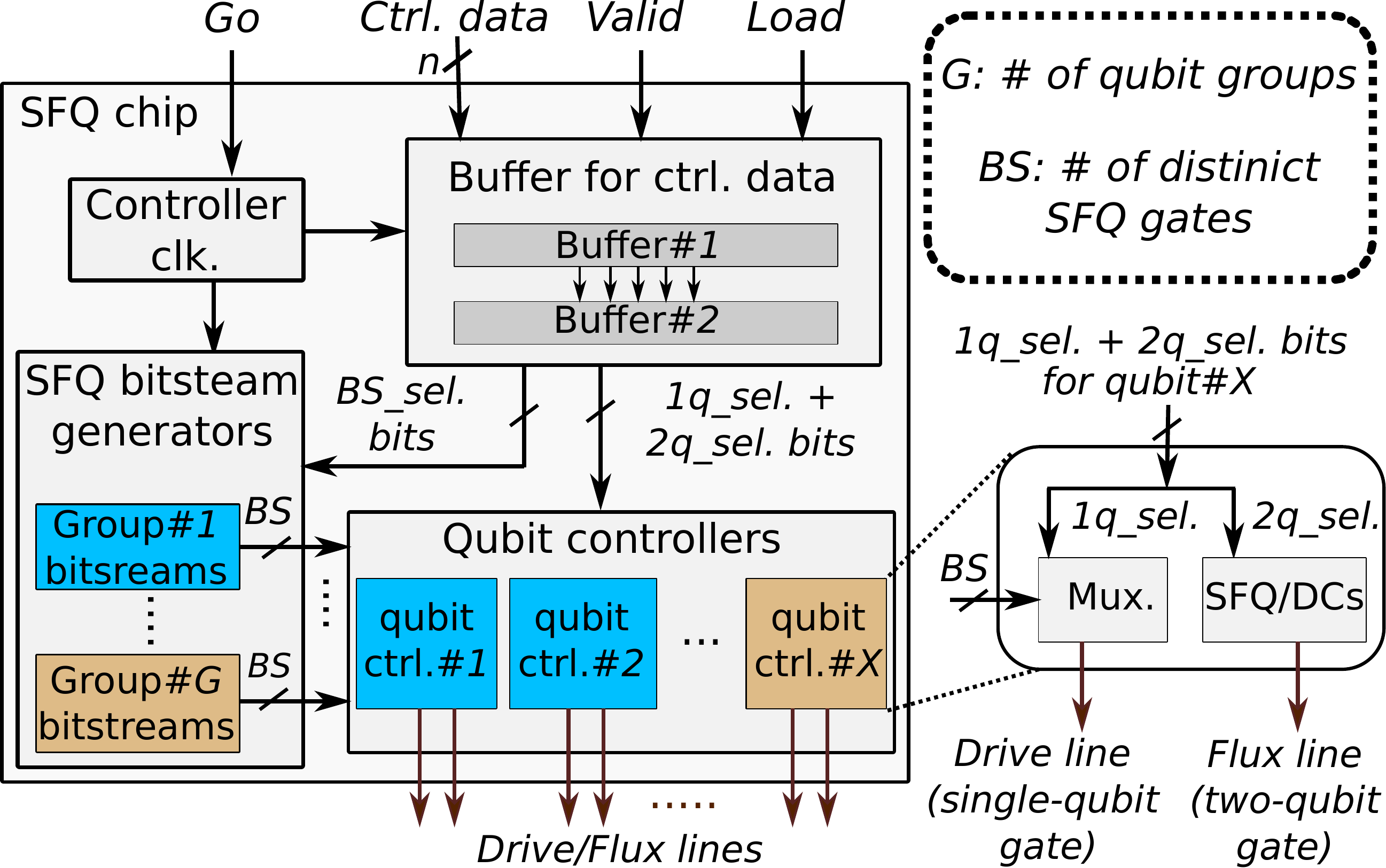}
  \caption{Overview of our \dq/ architecture.}
  \label{fig:overview}
  \end{center}
  \vspace{-17pt}
\end{figure}

\subsection{Overview of \dq/ architecture}
\label{sec:schemeoverview}
Now we put it all together and demonstrate an overview of our \dq/ architecture that is shown in Fig. \ref{fig:overview}. There are \emph{G} groups of qubit controllers. At the beginning of each controller cycle, \emph{BS} distinct single-qubit SFQ gates per each group are broadcasted to all the qubit controllers in the group. \emph{BS\_sel} bits are used to select the \emph{BS} distinct single-qubit gates in \dqopt/
(\dqmin/ does not need \emph{BS\_sel} bits since its universal single-qubit gate set includes only a few gates, which are all broadcasted).
Each qubit controller uses an SFQ-based multiplexer and the \emph{1q\_sel} bits to either select/apply one of the \emph{BS} delayed bitstreams, or apply none of them (e.g., in the two-qubit gate case). For two-qubit gates, the qubit controllers of the two qubits involved use the \emph{2q\_sel} bits to determine whether to start/stop performing CZ gate by sending \emph{start/stop} signals to the SFQ/DCs. 

\emph{BS\_sel}, \emph{1q\_sel}, and \emph{2q\_sel} control bits are sent from the room temperature every controller cycle using \emph{Ctrl. data} cables; a \emph{Valid} cable is used to determine the validity of control data on data cables. A \emph{Load} cable is also used to load the SFQ bitstreams through the data cables, which is done offline (i.e., not during the program execution); each SFQ bitstream has $\le$300 bits (the actual bitstream length depends on the target gate and system Hamiltonian). After the transmission of the control bits of the first controller cycle is finished, a \emph{Go} signal is sent from room temperature to start the controller clock (which is implemented using a counter that counts up every SFQ chip cycle and resets every controller cycle).
At the beginning of every controller cycle, 
the control bits that are already buffered in a buffer (\emph{Buffer\#1} in Fig. \ref{fig:overview}) are transferred to a second buffer (\emph{Buffer\#2} in Fig. \ref{fig:overview}) to be used by qubit controllers and SFQ bitstream generators. While executing the current controller cycle, the control bits of the next controller cycle are buffered in the first buffer.

\section{Software calibration of SIMD hardware}
\label{sec:simd}

\label{sec:drift-performance}

Real superconducting qubits exhibit variations in their oscillation frequency which can drift from day to day in experimental settings \cite{pranav,qarch_swamit3}.
Quantum gates are implemented using precise control signals designed using careful physical models of qubit evolution (that is, their Hamiltonian)~\cite{quantum_eng}, which are extremely sensitive to small deviations in qubit frequency.
The accuracy of quantum gates optimized for one set of qubit frequencies will therefore degrade rapidly if the same control signals are used on qubits exhibiting slight deviations from those frequencies.
State-of-the-art quantum gates therefore require regular recalibration using experimental measurements of each qubit \cite{pranav}.
For small systems with MIMD control, qubit-specific gate calibration can be performed at the hardware level by precisely shaping control pulses for each qubit. This is not possible for a SIMD architecture such as \dq/, in which control signals are shared among many qubits and so cannot be calibrated independently for each qubit.

\begin{figure}[t]
  \begin{center}
  \includegraphics[width=0.9\linewidth]{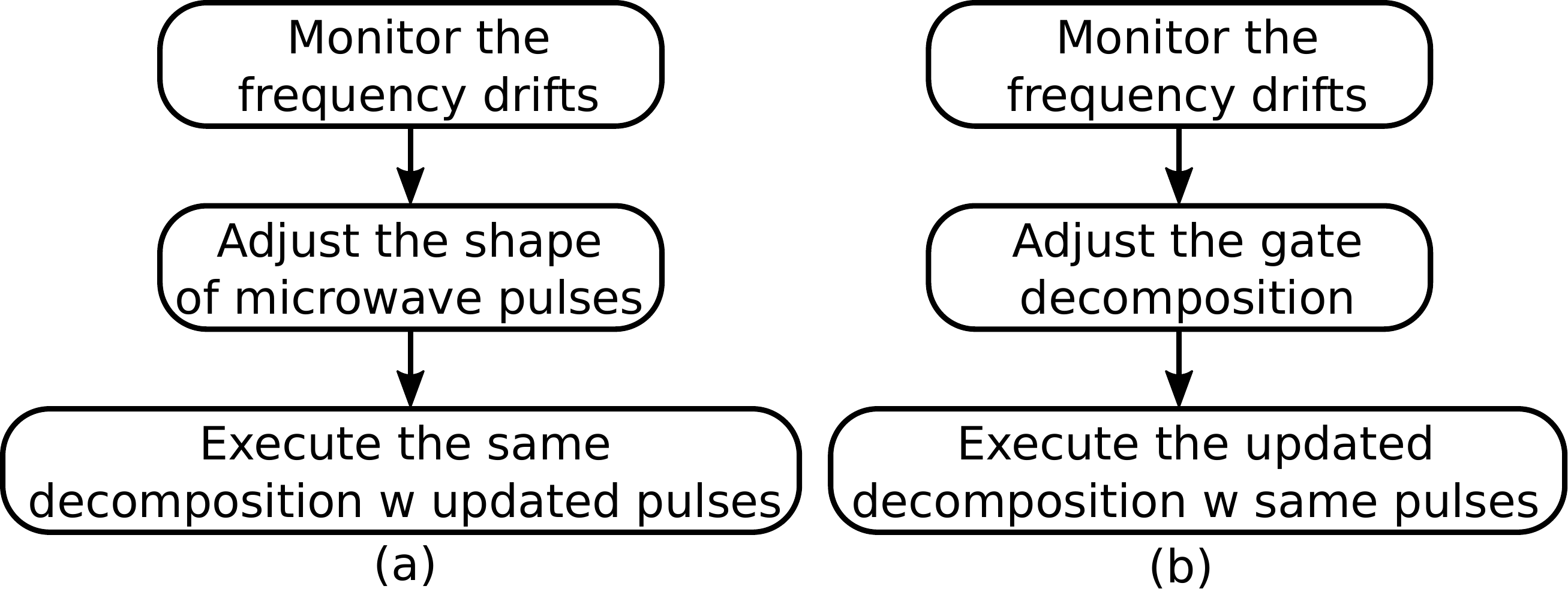}
  \caption{Calibration process in (a) today's microwave-based quantum machines; (b) \dq/.}
  \label{fig:calibration}
  \end{center}
  \vspace{-15pt}
\end{figure}

Here, we show that this challenge can be overcome by moving gate calibration to the software level, eliding unscalable qubit-specific hardware configurability.
Our approach is summarized in \cref{fig:calibration}. In short,
basis SFQ bitstreams that are stored on SFQ chip are fixed, and calibration is performed by adjusting the decomposition into the (frequency-dependent) set of basis quantum operations resulting from those fixed bitstreams; note that applying the same SFQ bitstream to multiple qubits with different qubit frequencies results in different quantum operations.
\revc{Though the focus of this section is compensation for frequency drift, similar techniques could be employed to mitigate other sources of systematic error.}

The gate errors reported through the remainder of this paper are calculated as $\epsilon=1-\overline{F}$, where $\overline{F}$ is the average gate fidelity \cite{avgfidelity,nufidelity}. We calculate gate fidelities by modeling and simulating the Hamiltonian of the quantum hardware to provide an upper bound for the gate fidelity.
This approach is also used in studies on both SFQ-based and microwave-based systems as a precursor to experimental work \cite{mcdermott_kangbo,sgrape}; \dq/ shows comparable gate fidelity to these studies as shown in Sec. \ref{sec:success-prob}.
Although we model the key dominant sources of systematic error, experimental evaluations on real hardware (currently available only at a small scale) are subject to various sources of noise which are not captured in the models, resulting in lower gate fidelities \cite{mcdermott_fab,echo2}.

\subsection{Calibrating single-qubit gates}
\label{sec:simd:1qb}

The key idea behind gate calibration on \dq/ is that these frequency-dependent operations still constitute universal gate sets for single-qubit computation \cite{kitaev1997}. We can therefore account for frequency drifts and hardware variation on each qubit by decomposing single-qubit gates on each qubit using the unique set of basis operations resulting from shared SFQ bitstreams.
For both \dqopt/ and \dqmin/, the single-qubit calibration procedure then consists of four steps:
(1) find SFQ bitstreams implementing a desired set of basis gates with high fidelity on qubits with no frequency variation,
(2) characterize each qubit's actual oscillation frequency using experimental measurements \cite{mcdermott_full},
(3) use the learned bitstreams and measured frequencies to determine the actual basis operations implemented on each qubit by the shared bitstreams, and
(4) compile quantum circuits using the actual single-qubit basis operations determined for each qubit.
The (classical) complexity of these steps scales linearly with number of qubits and circuit size.

For steps (1) and (3) we employ single-qubit physical simulations of the expected single-qubit unitary evolution $\ubs$ produced by an SFQ bitstream on transmon qubits (as done in \cite{mcdermott_kangbo}).
To ensure we are fully accounting for state leakage, we model transmons using six energy levels, and then compute single-qubit gate fidelities by projecting the resulting evolution back into the two-level subspace (causing any leakage to be captured as additional error \cite{nufidelity}).

For \dqopt/, calibrating gate decomposition means finding a new set of delay intervals such that the target operation is approximated by a sequence $\rz(\phi_{L})\ubs...\rz(\phi_{1})\ubs\rz(\phi_0)$ (where the final $\rz(\phi_{L})$ is absorbed into a subsequent gate).
We choose sets of delays holistically for each gate, numerically searching for the best combination of the available delays to implement that gate.
We find that $L\le2$ is sufficient for most gates, but a subset of gates nearing $\pi$ rotations around the x or y axis (e.g., Pauli X and Y operations) need $L=3$ to obtain high fidelity on qubits with the most significant drift (whereas in the ideal case (i.e., $\ubs=\yy$), $L\le2$ is enough for all single-qubit gates).

\begin{table}[!bt]
\caption{Optimal parking frequencies and drift tolerance for $\rz(\phi)$ gates with $\le10^{-4}$ error for $N=255$.}
\centering
\small
\begin{tabular}{|c|c|}
\hline
    Parking frequency (GHz) & Drift (GHz) for error $\le10^{-4}$ \\
\hline
     6.21286 & $\pm 0.01282$ \\
     5.02978 & $\pm 0.01049$ \\
     4.14238 & $\pm 0.00820$ \\
\hline
\end{tabular}
\label{tab:opt-freqs}
\vspace{-10pt}
\end{table}

Two factors affect performance of the \dqopt/ decomposition.
First, the set of available $\rz(\phi)$ rotations
is highly dependent on qubit frequency, which determines how well the $N+1$ possible delay values cover the unit circle. In the ideal case, the $N+1$ phases are equally spaced around the unit circle, and any $\rz(\phi)$ can be approximated to within $\pi/(N+1)$ radians; in this case we find that $N=255$ is sufficient for error $\epsilon\le0.25\cdot10^{-4}$.
We choose target frequencies with the highest tolerance for variation, as measured by the width of the interval in which any $\phi$ can be approximated with $<10^{-4}$ error using one of the available delays. These optimal parking frequencies and their drift tolerance are shown for $N=255$ in \cref{tab:opt-freqs}.
Second, frequency variations can limit SIMD scheduling.
For example, if a circuit calls for the same gate to be applied to many qubits, these may still require unique delay values when the decomposition of that gate differs for each qubit.
However, we can increase parallelism by allowing a small error margin when choosing delay values: often, multiple sets of delays will approximate the same operation with nearly equal error, so we can choose the one with lowest cost in terms of serialization.

For \dqmin/, we decompose arbitrary single-qubit gates into sequences of discrete, qubit-specific basis operations.
We use a brute-force search algorithm to find the optimal decomposition of single-qubit gates for each qubit, working with the full six-level representation of the unitary basis operations so that the decomposition accounts for and mitigates leakage resulting from each basis operation.

\subsection{Calibrating two-qubit gates}
\label{sec:simd:2qb}

The implementation of CZ gates using flux-tunable transmons requires the shape and duration of each current pulse to be carefully calibrated to the precise qubit frequencies and hardware parameters.
On a small system with MIMD control, these pulses can be individually calibrated for each pair of coupled qubits to account for variation and drift.
Without this fine-grained control, we are instead left with a unique 2-qubit operation $\uqq$ for each coupled pair of qubits. 
Here, we argue that we can instead compose CZ gates for each pair of qubits in \dq/ using pair-specific sequences of $\uqq$ operations and single-qubit gates, again relegating calibration to software.

We can compute the unitary evolution $\uqq$ for a pair of qubits using physical simulations of the empirical current waveform described in \cref{sec:scheme:2qubit}.
Starting with the nominal pre-drift frequencies of 6.21286 and 4.14238 GHz from \cref{tab:opt-freqs}, we vary each qubit's frequency and compare the resulting $\uqq$ to the target CZ operation.
We compute unitary evolution by numerically integrating the Schr\"odinger equation using well-understood Hamiltonian models of pairs of capacitively-coupled flux-tunable asymmetric transmons \cite{quantum_eng}.
We assume that each transmon has an anharmonicity of 250 MHz, 
and the capacitive coupling strength is 10 MHz.

\begin{figure}[!t]
  \centering
    \includegraphics[ width=\linewidth]{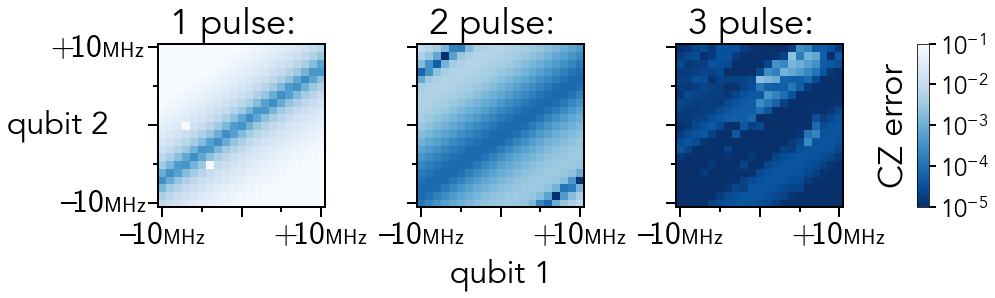} 
    \\\vspace{-2pt}
    \;\;(a)\qquad\qquad\;\quad(b)\qquad\qquad\quad(c)\qquad
  \caption{CZ gate error as a function of frequency drift, assuming 1, 2, or 3 $\uqq$ operations and ideal single-qubit gates.
  }
  \label{fig:cz:freq-shift}
\end{figure}

In \cref{fig:cz:freq-shift}(a), we show the expected minimum error when implementing a CZ gate using a single $\uqq$ pulse as a function of each qubit's drift, allowing for arbitrary single-qubit gates before and after the pulse.
At the ideal qubit frequencies, we find an expected error of $\epsilon=3\cdot10^{-4}$.
This error grows rapidly as the frequency difference between qubits drifts.
In \cref{fig:cz:freq-shift}(b) and (c), we show the gate error after compiling CZ gates into sequences of 2 or 3 $\uqq$ operations and intermediary single-qubit gates, similar to the ``echo'' sequences described in \cite{echo1,echo2} but with single-qubit gates obtained via numerical optimization.
Assuming ideal single-qubit gates, we find that 3 $\uqq$ operations are sufficient to achieve $\epsilon\le10^{-4}$ over a wide range of frequencies.
In \cref{sec:success-prob}, we report empirical gate errors after single-qubit gates are decomposed for either \dqopt/ or \dqmin/.

\section{Methodology and Results}
\label{sec:results}
 \label{sec:sfq_cells}
\begin{table}[!bt]
\small
\centering
\caption{The library of RSFQ cells and corresponding characteristics used for synthesis.
}
\begin{tabular}{@{\extracolsep{4pt}}|l|c|c|c|@{}}
\hline
Cell & Area ($\mu$m$^2$) & JJ Count & Delay (ps)\\ \hline
AND2 & $3500$ & $16$ & $8.4$\\
OR2 & $3500$ & $14$ & $6.1$\\
XOR2 & $3500$ & $18$ & $5.8$\\
NOT & $3500$ & $12$ & $13.2$\\
DRO DFF & $3000$ & $11$ & $6.2$\\
NDRO DFF & $4500$ & $18$ & $9.3$\\
Splitter & $2000$ & $6$ & $7.1$\\
\hline
\end{tabular}
\label{table:cells}
\vspace{-10pt}
\end{table}

In this section, we present hardware synthesis results of \dq/, followed by
an analysis of its algorithmic performance.

\subsection{Hardware results of \dq/}
We use detailed SFQ synthesis tools \cite{pasandi2019balanced, pasandi2018sfqmap, pasandi2019pbmap, pasandi2019dynamic, pasandi2019efficient} to synthesize, map, and finally calculate the power, area, and delay values. The employed synthesis and technology mapping flow is as follows: first, technology-independent optimizations including balanced factorization and rewriting \cite{pasandi2019balanced} are performed on the input circuit, then the circuit is mapped using a path balancing technology mapping algorithm \cite{pasandi2019dynamic} and fully path balanced \cite{pasandi2019pbmap}. Next, a standard retiming algorithm similar to \cite{leiserson1991retiming} is employed to further reduce the total memory element count. Finally, the memory elements (e.g., latches) are replaced with SFQ DRO DFFs, and splitters are inserted at the output of gates with more than one fanout. 

Rapid Single Flux Quantum (RSFQ) logic family \cite{sfq_logic2} is used in this paper, and the library of cells is listed in Table \ref{table:cells}. 
For wiring, we use Josephson Transmission Line (JTL) and Passive Transmission Line (PTL). JTL is used for short connection for cells next to each other and its delay is $\sim$1.5-2 ps.
PTL is used for transmitting SFQ pulses from one logic gate to another. 

\subsubsection{Validity of our synthesis results}
We have the physical layouts for all cells, including wires and logic cells. Our RSFQ power, area and delay numbers are obtained post-layout based on accurate extraction of the gate layouts and passive transmission lines and subsequently simulated using WRSpice \cite{wrspice} and JSim \cite{jsim}. The simulation results are thus highly accurate and reliable. The SFQ library cells have been validated and their power/timing values calibrated against manufactured test chips done in the MIT Lincoln lab’s SFQ5ee process node \cite{sfq5ee}. The synthesis, place\&route, modeling/simulation, and formal verification tools have been used to design and formally/simulatively verify tens of SFQ circuits ranging from individual data path modules and register files to major building blocks of the RISC-V Sodor core \cite{haolin,coldflux,sodor}. In addition, prior work validated comparable SFQ tools and simulators with SFQ chip fabrication \cite{supernpu}.

\subsubsection{Delay results}
Our synthesis results show that the worst stage delay in \dq/ is 34.5 ps, which determines the maximum clock frequency that our SFQ chip can work with. We choose 40 ps as our SFQ chip clock period (similar to \cite{mcdermott_kangbo}), thus the bit period in our SFQ bitstreams is 40 ps.

\captionsetup[subfigure]{labelformat=empty}
\begin{figure}[!t]
\begin{center}
\subfloat    {
\includegraphics[width=0.92\linewidth]{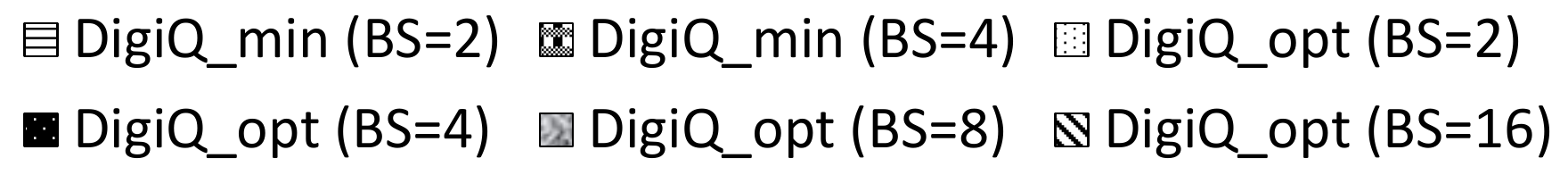}
}\\
\subfloat [(a)]    {
\includegraphics[width=0.87\linewidth]{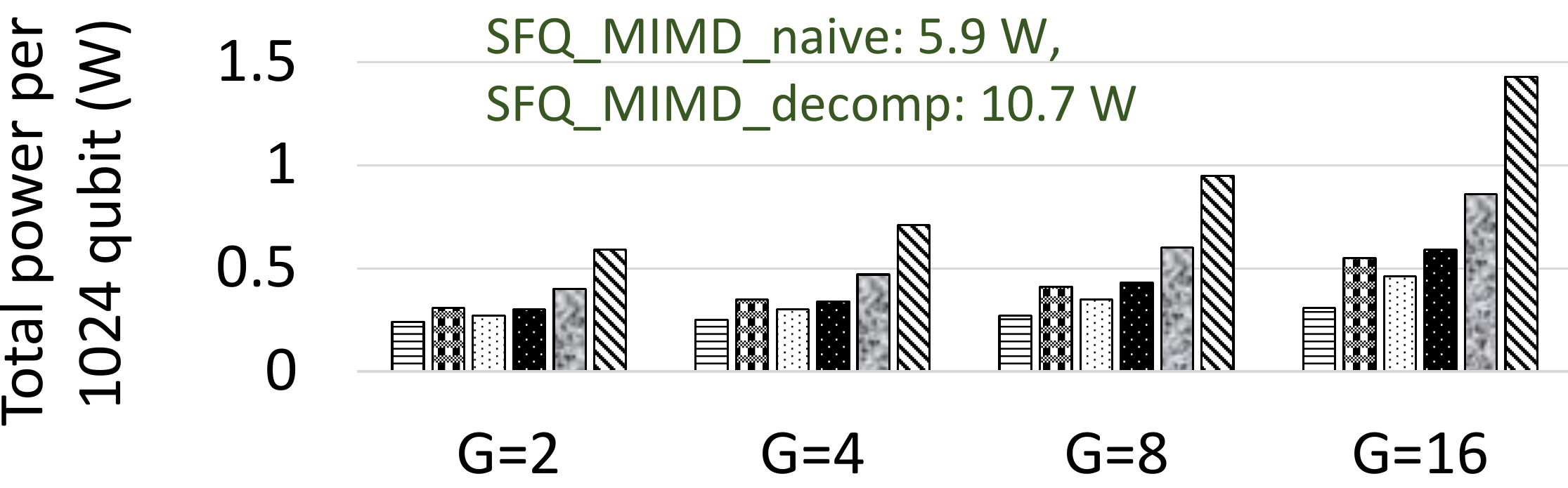}
} \\\vspace{-5pt}
\subfloat  [(b)]   {
\includegraphics[width=0.87\linewidth]{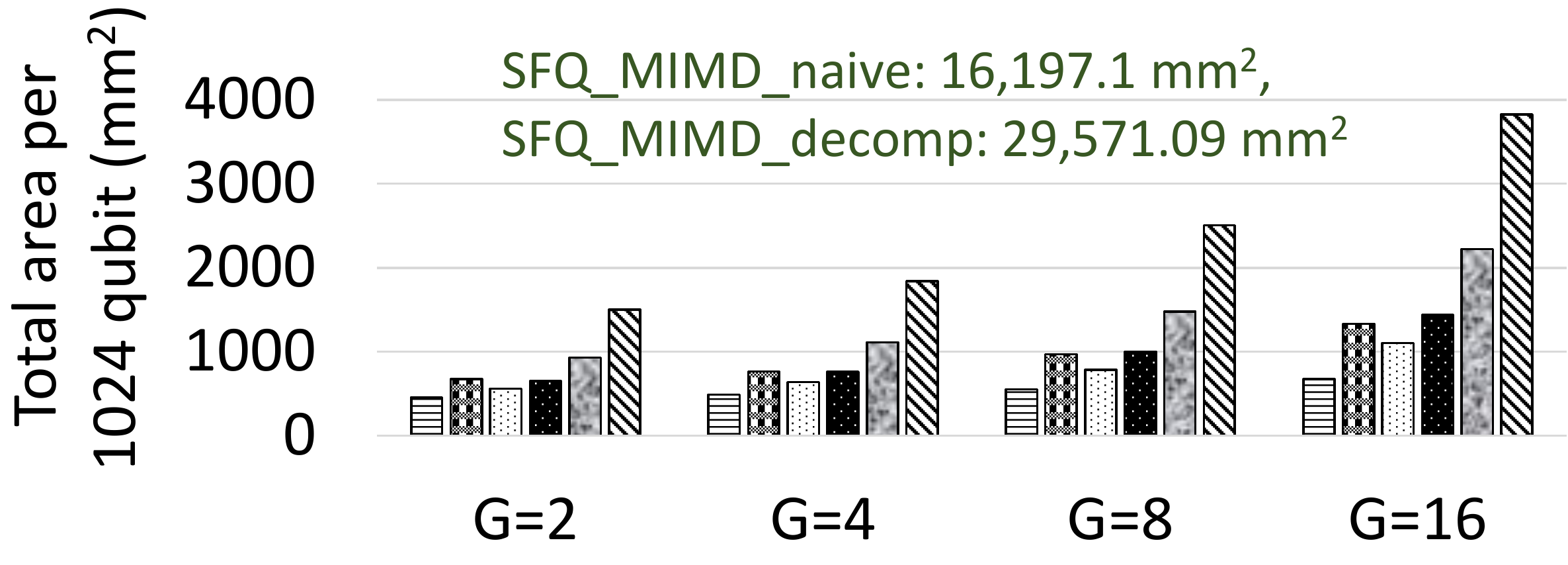}
} \\\vspace{-5pt}
\subfloat [(c)]    {
\includegraphics[width=0.87\linewidth]{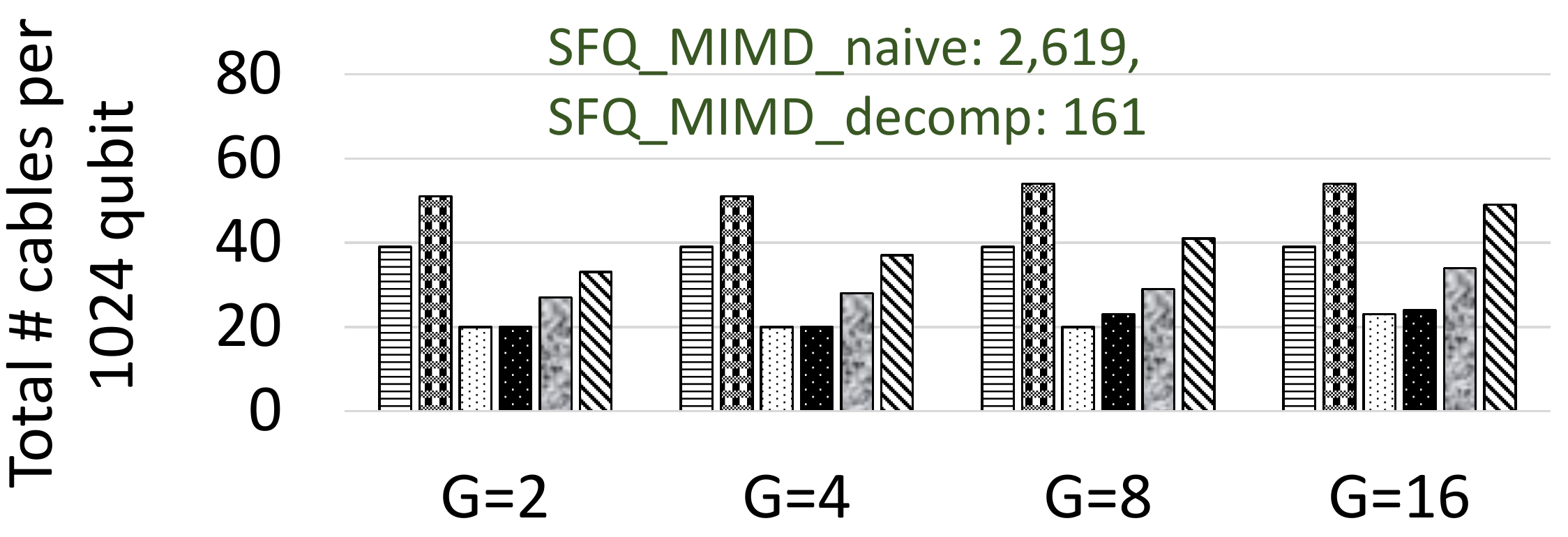}
}
\caption{Power (a), area (b), and cable count (c) results of \dqmin/ and \dqopt/ architectures. \revc{\sfqnaive/ and \sfqmimd/ results are shown for comparison.}}
\vspace{-25pt}
\label{fig:area_pow}
\end{center}
\end{figure}

\subsubsection{Power and area results} 
\label{sec:pow}
Fig. \ref{fig:area_pow}(a) and Fig. \ref{fig:area_pow}(b) show the total power and area of \dq/, respectively. We present the results for different $BS$ and $G$ values.

As mentioned in Sec. \ref{sec:scheme:one-qubit}, in \dqopt/, only $BS$ distinct delays are available every controller cycle for single-qubit gates, and that can potentially lead to quantum gate serialization inside a group. One solution to mitigate this serialization is to increase the value of $BS$, which would increase the hardware cost of quantum controller modules as they would need more complex SFQ-based multiplexers to select one out of a larger number of delayed bitstreams (see Fig. \ref{fig:overview}). Another solution to mitigate the serialization is to increase the $G$ value. This solution decreases the number of qubits per each group which means less congestion and less serialization, while not increasing the hardware cost of gate controller modules. Thus, at the design time, we expected that among all the designs with the same $BS*G$, the ones with higher $G$ value have lower hardware cost as they need less complex quantum controller modules. However, our synthesis results surprised us. As shown in Fig. \ref{fig:area_pow}, the hardware cost of the designs with the same $BS*G$ value does not differ significantly. The reason is that increasing the $G$ value also increases the overall hardware cost due to an increase in the number of SFQ bitstream generators (see Fig. \ref{fig:overview}). Given our synthesis results, we conclude that keeping the $G$ value small and increasing the $BS$ value is preferred because it provides more flexibility in allocating delayed bitstreams to qubits.
Smaller $G$ values are preferred for \dqmin/ as well, as increasing the $G$ value increases the hardware cost with no significant algorithmic advantage.
The smallest $G$ value or the maximum number of qubits in a group is determined by the area of one SFQ chip; if we cannot fit a group on one chip, we need to use multiple chips and replicate the bitstreams on each chip to avoid long distance communications, which is equivalent to dividing the qubits into multiple groups. Our results show that for 1024-qubit scale, we can fit all the designs with $G=2$ on at most two SFQ chips (one per group), thus we use $G=2$ for 1024-qubit benchmarks in Sec. \ref{sec:algorithms}.

Our results show that even our largest designs can operate within the power budget of typical dilution refrigerators at 4 K stage (i.e., a few watts \cite{mcdermott_full,qarch_hornibrook,horseridge2}), and can be implemented on only a few SFQ chips at 1,024-qubit scale; clock synchronization between the SFQ chips is done using SFQ phase locked loops (PLLs) \cite{sfqpll}. We replicate the 1,024-qubit design in order to scale to larger number of qubits (note that replicating the 1,024-qubit design naturally increases the number of groups). \dqmin/($BS$=$2$) has the lowest hardware cost and highest scalability ($>$42,000 qubits given 10 W power budget \cite{mcdermott_full}). Increasing the $BS$ value in \dqmin/ increases the hardware cost, but also can potentially increase the algorithmic performance
(we see diminishing returns after $BS=4$). The scalability of \dqopt/ is also high, allowing $>$25,000 qubits ($>$17,000 qubits) for $BS=8$ ($BS=16$).

\subsubsection{SFQ storage and Cable count results}
\dqmin/ stores $BS$ bitstreams per group and each bitstream has $\le$300 bits. 
\dqopt/ stores one bitstream with $\le$300 bits 
per group which is delayed by $BS$ different values at each controller cycle (see Sec. \ref{sec:scheme}). With both designs, for each qubit at each controller cycle, we need enough control bits from room temperature to determine whether to apply one of the $BS$ distinct gates, start/stop performing two-qubit gates, or perform no operation.
For \dqopt/, an additional $BS$ delay values for each group at each cycle are needed; since we have 256 possible delay values, each delay value requires $\log_{2} {256}=8$ bits.

Fig. \ref{fig:area_pow}(c) shows the number of required cables to send \emph{Go}, \emph{Valid}, \emph{Load}, \emph{BS\_sel}, \emph{1q\_sel}, and \emph{2q\_sel} bits from the room temperature in one controller cycle given 10 Gbps return-to-zero (RZ) cables \cite{sfq_logic2}.
We calculate the number of data cables assuming a minimum controller clock period of 9 ns for \dqmin/
and 9 ns + 10.2 ns for \dqopt/ (10.2 ns corresponds to 255 delay cycles) -- we need enough data cables to send one set of control bits from the room temperature within one controller cycle (see Sec. \ref{sec:schemeoverview}). \dqmin/($G$=2,$BS$=2) requires only 39 cables per 1,024 qubits (52.5X less than a microwave-based quantum computer with 2 cables per qubit, 1 drive line and 1 flux line \cite{google_machine1}). \dqopt/ requires just 33 cables per 1,024 qubits in $G=2$, $BS=16$ design.
\dqmin/'s high cable count relative to most \dqopt/ configurations is due to its shorter controller cycle.

\subsection{Algorithmic performance results of \dq/}
\label{sec:algorithms}
In order to study the algorithmic impacts of \dq/, we compile a common set of NISQ benchmarks for each design. The benchmarks chosen are described in \cref{tab:comp:benchmarks}, and  together represent a diverse sample of common circuit formulations.
Benchmark circuits are algorithmically generated and mapped to a $32\times32$ square grid via $\swap$-gate insertion using the stochastic transpiler pass packaged with Qiskit Terra~\cite{qiskit}. 
Each circuit is then decomposed into CZ and single-qubit gates,
and a crosstalk-aware scheduling pass~\cite{murali2020} is used to sort and group commuting two-qubit gates which can be executed simultaneously without interference.

\begin{table}[t!]
\caption{NISQ benchmark algorithms.}
\centering
\small
\begin{tabular}{rl}
\hline
\textbf{QGAN } & Quantum generative adversarial learning network~\cite{lloyd2018}\\
\textbf{Ising} & Linear Ising model spin chain simulation~\cite{Barends2016} \\ 
\textbf{BV   } & 1024-bit Bernstein-Varzirani algorithm~\cite{Bernstein1997} \\ 
\textbf{Add1} & 256-bit ripple-carry adder~\cite{Cuccaro2004} \\ 
\textbf{Add2} & 256-bit parallel carry-lookahead adder~\cite{Draper2006} \\
\textbf{Sqrt10} & 10-bit square root via Grover search~\cite{JavadiAbhari,Prakash2019,grover}\\
\hline
\label{tab:comp:benchmarks}
\end{tabular}
\vspace{-20pt}
\end{table}

To model frequency variation, each qubit is modeled as an asymmetric transmon with $\sigma=0.2\%$ variability in each of its Josephson energies (sampled from a normal distribution). At our target frequencies, this corresponds to about $\pm6$ MHz fluctuation in oscillation frequency, in keeping with design targets for future superconducting systems \cite{hertzberg2020}.
Hardware variability is considered with the addition of a $\sigma=1\%$ error to the output of each current generator.
As in \cref{sec:simd}, we use these sampled hardware parameters to physically simulate each basis operation on each qubit or qubit pair (similar to prior work \cite{sgrape,mcdermott_kangbo}). We then use the resulting unitary basis operations to decompose each gate in the benchmark circuits. Gate errors in \cref{sec:success-prob} are determined by computing the overlap between the decomposed and target gates. 

For \dqopt/, we use a 20.32 ns controller cycle time, comprising 10.12 ns for SFQ bitstreams and 255 delay cycles.
The CZ gate time is 60 ns (determined from the analysis in \cref{sec:simd:2qb}), which expands over three controller cycles.
Single-qubit gates are expanded into at most three $\ubs$ pulses (see Sec. \ref{sec:simd:1qb}).
For \dqmin/, single-qubit gate times for the 6.21286 and 4.14238 GHz qubit frequencies are respectively 10.12 ns and 9.00 ns, again with a 60 ns CZ gate time. We decompose single-qubit gates until the approximation error falls below $10^{-4}$, up to a maximum depth of 28 (at which point we find that only about 1\% of gates have errors above $10^{-4}$).
There is no feedback loop in our benchmarks, thus we do not consider the communication latency from outside the fridge.

\begin{figure}[t]
  \centering
  \subfloat[]      {
    \includegraphics[ width=0.9\linewidth]{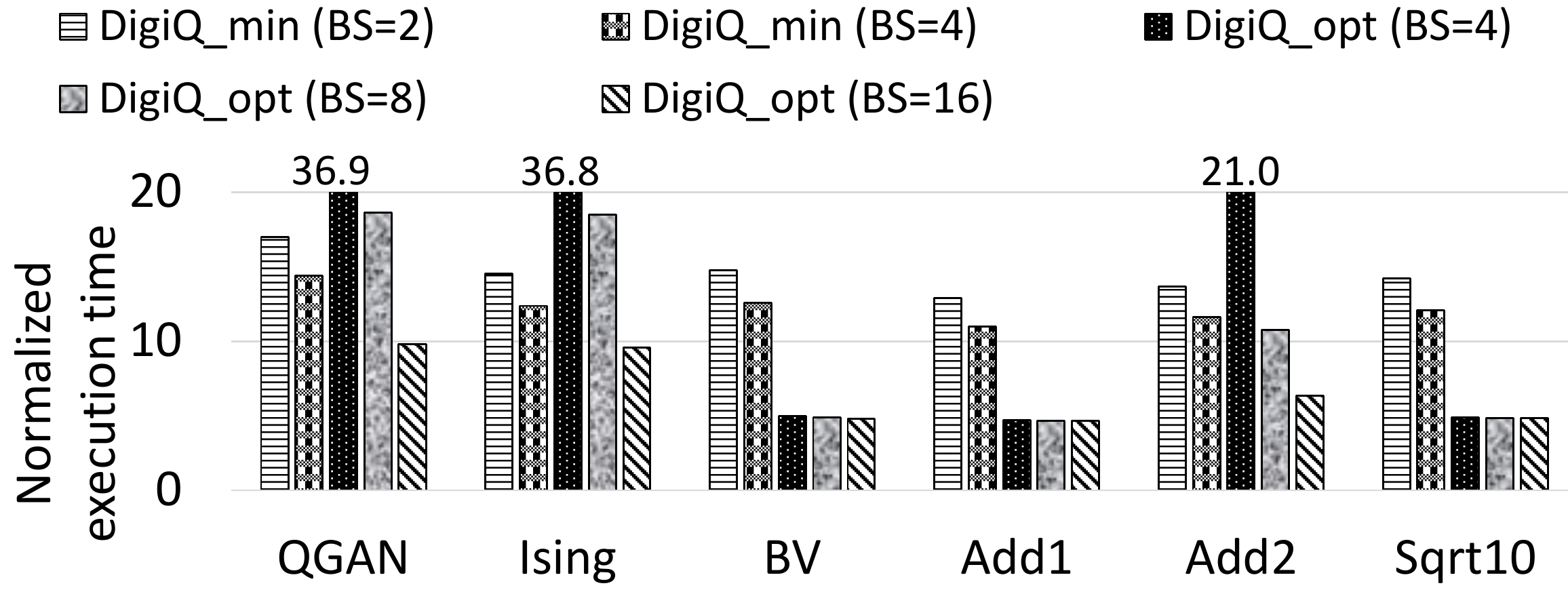}
      \vspace{-4pt}
    }
  \begin{center}
  \vspace{-20pt}
  \caption{\dq/ quantum circuit execution time normalized to the Impossible MIMD system.
  }
  \label{fig:comp:depths}
  \end{center}
  \vspace{-25pt}
\end{figure}

\subsubsection{Circuit execution time}
\label{sec:circuitexe}

Fig. \ref{fig:comp:depths} shows the execution time results for \dq/ with 1024 qubits, normalized to an \emph{Impossible MIMD} system which is assumed to have the same gate times as \dq/ (which are also similar to those of recent microwave-based systems \cite{google_machine1,pranav}) but with unlimited parallelism.
We emphasize that the MIMD system is impossible at large scales (see Sec. \ref{sec:back}); we compare our results with the Impossible MIMD system just to quantify the serialization cost of realizing a controller design that is feasible at large scales, and to
give readers a sense of how \dq/ would compare to today's prototypes if they did not have scalability issues.
Both \dqopt/ and \dqmin/ have some overhead in terms of execution time compared to the Impossible MIMD system (\dqopt/ due to serialization and \dqmin/ due to longer gate decompositions).
The performance of \dqopt/ increases by increasing the $BS$ value, as expected, with $BS=16$ providing the best performance.
This difference is minimal for benchmarks without significant parallelism (BV, Add1, Sqrt10). 
\revc{For $BS=16$, the execution time is only $\le 2X$ longer than what we would be able to achieve if \dqopt/ could be built with zero quantum gate serialization (i.e., $BS=\infty$).}
In \dqmin/, increasing $BS$ from 2 to 4 reduces the depth of single-qubit gate decompositions by roughly half. However, the benefit is limited beyond $BS=4$ due to the dominance of CZ gate decompositions which do not benefit substantially from the larger single-qubit gate set. 
\dqmin/ performs similarly to \dqopt/($BS$=$8$) for the benchmarks with more parallelism (QGAN, Ising, Add2), in which \dqopt/ experiences more gate serialization. For the remaining benchmarks,
\dqmin/'s long gate decomposition leads to worse performance than any \dqopt/ configuration.
Compared to Impossible MIMD system, \dqopt/($BS$=$16$) is 4.7X-9.8X worse in terms of circuit execution time, while \dqmin/($BS$=4) is 11.0X-14.4X worse.

\captionsetup[subfigure]{labelformat=empty}
\begin{figure}[!t]
\begin{center}
\subfloat  [(a)]  {
\includegraphics[width=0.85\linewidth]{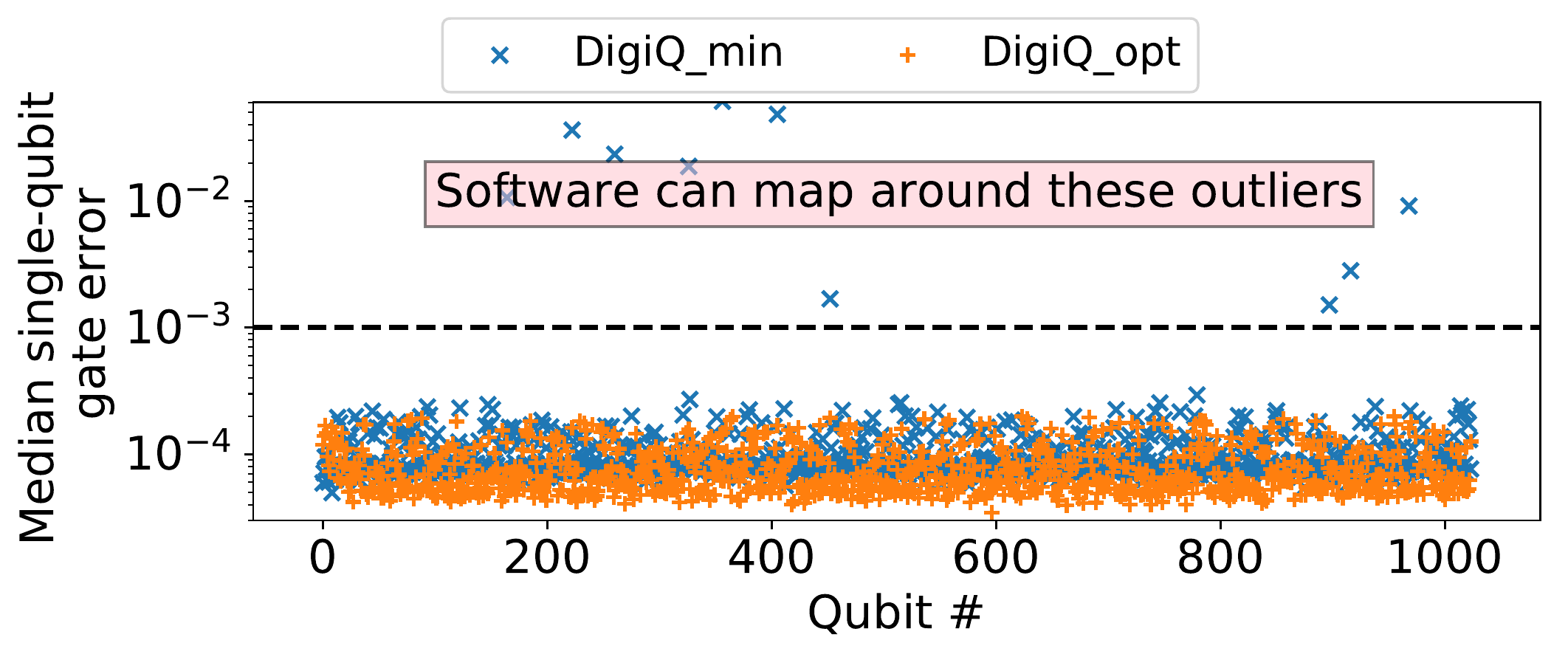}
}\\ \vspace{-10pt}
\subfloat [(b)]   {
\includegraphics[width=0.85\linewidth]{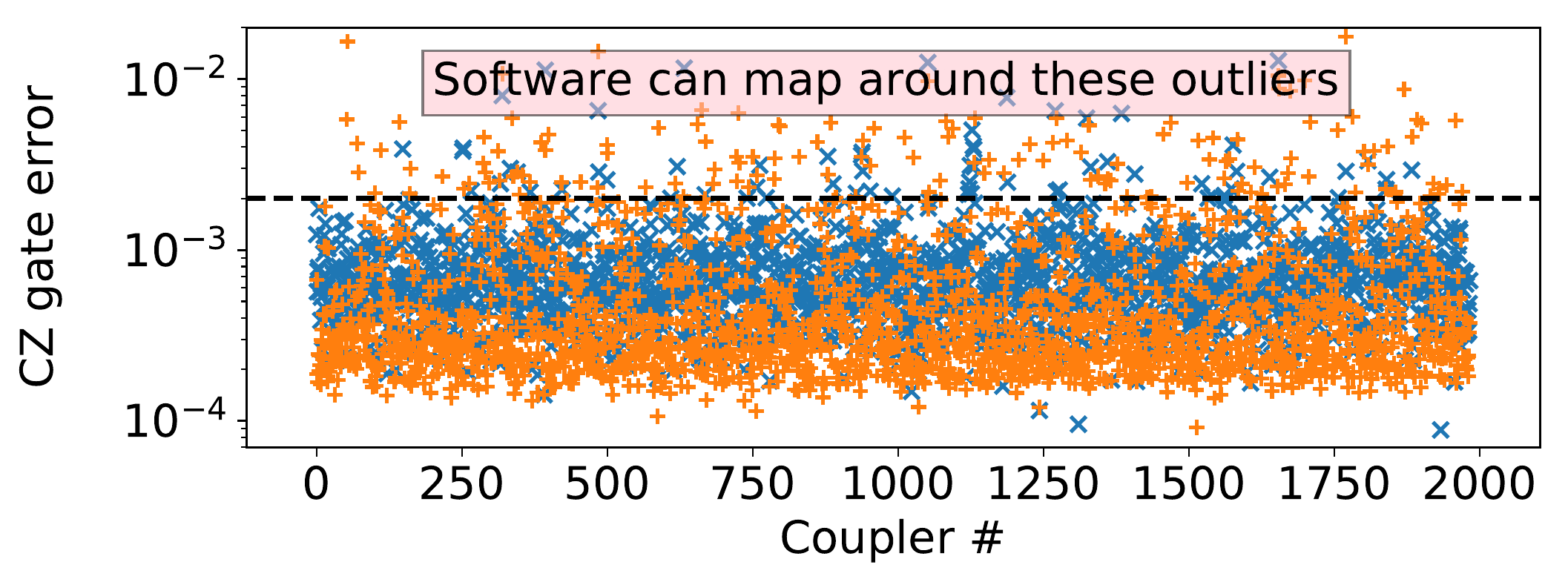}
} \vspace{-5pt}
\caption{(a) Median single-qubit gate error on \dqopt/ ($BS$=8) and \dqmin/ ($BS$=2) with 1024 qubits (well representative of other configurations); (b) CZ gate error on each qubit pair. Software can map around the outliers using the noise-adaptive mapping techniques \cite{prakash}.}
\vspace{-20pt}
\label{fig:gate-error}
\end{center}
\end{figure}

\subsubsection{Gate/circuit error}
\label{sec:success-prob}
 
We estimate the overall circuit error due to gate decomposition by taking the product of the errors of each of its gates.
For both designs, typical gate error varies between qubits with different frequency drifts.
\cref{fig:gate-error}(a) shows the median error of single-qubit gates (evaluated across all gates in all benchmarks) for each qubit on \dqopt/($BS$=8) and \dqmin/($BS$=2) (the same conclusions hold for other $BS$ values). We find that \dqopt/ exhibits higher variability between qubits, whereas \dqmin/ is generally more stable but has a small number of especially bad outlier qubits (where the frequency drift brings the nominal T gate very close to the identity, resulting in a poor single-qubit gate set).
Similar to microwave-based systems, we can map around these outlier cases in software using noise-adaptive mapping techniques \cite{prakash}.
The CZ gate error (\cref{fig:gate-error}(b)) is driven both by the decomposition of CZ into $\uqq$ and single-qubit gates, and by the error with which we can decompose those single-qubit gates on each architecture. The CZ error is $>$0.002 for 3\% and 7\% of qubit pairs in \dqmin/ and \dqopt/, respectively. Note that the CZ error would be $>$0.002 for 84\% of qubit pairs if we did not use our software calibration techniques.
Our results show that with software calibration \dq/ can achieve similar gate error to that of hardware-calibrated microwave-based gates \cite{sgrape}.

\section{Conclusions and Future work}
\label{sec:conclusion}

Large-scale quantum computers are essential in order to perform many useful quantum algorithms. However, superconducting quantum computer prototypes have severe scalability issues due to the massive overhead of generating and routing control pulses from room temperature to quantum chip inside the dilution refrigerator. In this work, we present \dq/, the first system-level design of a digital SFQ-based quantum controller architecture to maximize scalability. We provide architecture guidelines to design large-scale SFQ-based controllers by taking into consideration the opportunities (efficient on-chip broadcast and ultra-fast clock) and challenges (lack of dense memory/logic) of SFQ.
Our study shows that we must deploy a SIMD architecture to operate within the tight power/area budget of dilution refrigerators,
however, SIMD also introduces new challenges in terms of control pulse calibration in NISQ machines with imperfect qubits.
We propose novel software-solution to address these calibration challenges efficiently, and show that software plays a key role in ensuring good quantum algorithmic performance at large scales. We fully characterize \dq/ controller hardware using state-of-the-art/validated SFQ synthesis tools. We also show the effectiveness of \dq/ in terms of quantum algorithmic performance under a variety of NISQ algorithms. We model dominant sources of systematic error in our evaluations and show that \dq/ has similar fidelity to microwave-based systems under the same models. Our analysis shows that \dq/ is a realistic path forward to realize large scale quantum computers, and we hope the promising results of this paper motivate experimentalists to further explore SFQ-based controllers. 

Going forward, the fidelity of quantum gates/circuits can further be improved by performing further research at both hardware and software levels. First, the main bottleneck in increasing the gate fidelity in \dq/ is frequency drift of imperfect qubits, thus developing better qubits with less drift 
can increase the fidelity significantly. Second, decreasing the power and area consumption of SFQ circuits would allow the realization of more complex SFQ-based controllers that can potentially perform hardware calibration, which combined with our software calibration can further increase the fidelity of quantum gates/circuits.
Third, further research at the software level can potentially lead to more robust decompositions with higher fidelity. 

Finally, we would like to mention that the ideas proposed in this paper such as SIMD design and software calibration can potentially increase the scalability of today's Cryo-CMOS controllers as well. However, 
further research on such controllers based on Cryo-CMOS is needed. First, although SIMD can reduce the cost of on-chip storage in Cryo-CMOS,
novel approaches are needed to share/broadcast the instructions (i.e., amplitudes/phases of microwave pulses) on the Cryo-CMOS chip with low cost to obtain overall cost reduction. Second, Cryo-CMOS has denser memory/logic, but much higher power, which would result in significantly different set of design tradeoffs. Third, \dq/ relies on ultra-fast clock in SFQ (i.e., in the order of ps), and 
Cryo-CMOS requires different architectures and gate implementations as it works with clocks in the order of ns.

\section*{Acknowledgment}

This work is funded in part by EPiQC, an NSF Expedition in Computing, under grants CCF-1730082/1730449; in part by STAQ under grant NSF Phy-1818914; in part by DOE grants DE-SC0020289 and DE-SC0020331; and in part by NSF OMA-2016136 and the Q-NEXT DOE NQI Center. This work was completed in part with resources provided by the University of Chicago’s Research Computing Center.

\let\oldbib\thebibliography
\def\thebibliography#1{\oldbib{#1}\setlength{\itemsep}{5pt}}
\bibliographystyle{IEEEtran}
\bibliography{references-clean}

\begin{thebibliography}{10}
\providecommand{\url}[1]{#1}
\csname url@samestyle\endcsname
\providecommand{\newblock}{\relax}
\providecommand{\bibinfo}[2]{#2}
\providecommand{\BIBentrySTDinterwordspacing}{\spaceskip=0pt\relax}
\providecommand{\BIBentryALTinterwordstretchfactor}{4}
\providecommand{\BIBentryALTinterwordspacing}{\spaceskip=\fontdimen2\font plus
\BIBentryALTinterwordstretchfactor\fontdimen3\font minus
  \fontdimen4\font\relax}
\providecommand{\BIBforeignlanguage}[2]{{%
\expandafter\ifx\csname l@#1\endcsname\relax
\typeout{** WARNING: IEEEtran.bst: No hyphenation pattern has been}%
\typeout{** loaded for the language `#1'. Using the pattern for}%
\typeout{** the default language instead.}%
\else
\language=\csname l@#1\endcsname
\fi
#2}}
\providecommand{\BIBdecl}{\relax}
\BIBdecl

\bibitem{ibm_device}
M.~Brink, J.~M. Chow, J.~Hertzberg, E.~Magesan, and S.~Rosenblatt, ``Device
  challenges for near term superconducting quantum processors: frequency
  collisions,'' in \emph{Proc. IEEE Int. Electron Devices Meeting (IEDM)}, Dec.
  2018, pp. 6.1.1--6.1.3.

\bibitem{qarch_yuan}
G.~Li, Y.~Ding, and Y.~Xie, ``Towards efficient superconducting quantum
  processor architecture design,'' in \emph{Proc. 25th Int. Conf. Archit.
  Support Program. Lang. Oper. Syst. (ASPLOS)}, Mar. 2020, pp. 1031--1045.

\bibitem{google_machine1}
F.~Arute, K.~Arya, R.~Babbush, D.~Bacon \emph{et~al.}, ``Quantum supremacy
  using a programmable superconducting processor,'' \emph{Nature}, vol. 574,
  no. 7779, pp. 505--510, Oct. 2019.

\bibitem{google_machine2}
\BIBentryALTinterwordspacing
J.~Kelly, ``A preview of {Bristlecone}, {Google}’s new quantum processor,''
  \emph{Google Research Blog}, Mar. 2018. [Online]. Available:
  \url{https://ai.googleblog.com/2018/03/a-preview-of-bristlecone-googles-new.html}
\BIBentrySTDinterwordspacing

\bibitem{ibm_machine}
M.~Steffen, D.~P. {DiVincenzo}, J.~M. Chow, T.~N. Theis, and M.~B. Ketchen,
  ``Quantum computing: an {IBM} perspective,'' \emph{IBM Journal of Research
  and Development}, vol.~55, no.~5, pp. 13:1--13:11, 2011.

\bibitem{qarch_fu}
X.~Fu, M.~A. Rol, C.~C. Bultink, J.~van Someren \emph{et~al.}, ``An
  experimental microarchitecture for a superconducting quantum processor,'' in
  \emph{Proc. 50th Annu. IEEE/ACM Int. Symp. Microarchit. (MICRO)}, Oct. 2017,
  pp. 813--825.

\bibitem{mcdermott_full}
R.~{McDermott}, M.~Vavilov, B.~Plourde, F.~Wilhelm \emph{et~al.},
  ``Quantum--classical interface based on single flux quantum digital logic,''
  \emph{Quantum Science and Technology}, vol.~3, no.~2, 2018.

\bibitem{mcdermott_fab}
E.~Leonard, M.~A. Beck, J.~Nelson, B.~Christensen \emph{et~al.}, ``Digital
  coherent control of a superconducting qubit,'' \emph{Phys. Rev. Applied},
  vol.~11, no.~1, Jan. 2019.

\bibitem{mcdermott_kangbo}
K.~Li, R.~McDermott, and M.~G. Vavilov, ``Hardware-efficient qubit control with
  single-flux-quantum pulse sequences,'' \emph{Phys. Rev. Applied}, vol.~12,
  no.~1, Jul. 2019.

\bibitem{horseridge2}
J.~P.~G. Van~Dijk, B.~Patra, S.~Subramanian, X.~Xue \emph{et~al.}, ``A scalable
  cryo-{CMOS} controller for the wideband frequency-multiplexed control of spin
  qubits and transmons,'' \emph{{IEEE} J. Solid-State Circuits}, vol.~55,
  no.~11, pp. 2930--2946, 2020.

\bibitem{sfq_logic1}
D.~Kirichenko, S.~Sarwana, and A.~Kirichenko, ``Zero static power dissipation
  biasing of {RSFQ} circuits,'' \emph{{IEEE} Trans. Appl. Supercond.}, vol.~21,
  no.~3, pp. 776--779, 2011.

\bibitem{sfq_logic2}
K.~K. Likharev and V.~K. Semenov, ``{RSFQ} logic/memory family: A new
  {Josephson}-junction technology for sub-terahertz-clock-frequency digital
  systems,'' \emph{{IEEE} Trans. Appl. Supercond.}, vol.~1, no.~1, pp. 3--28,
  1991.

\bibitem{sfq_genetic}
P.~J. Liebermann and F.~K. Wilhelm, ``Optimal qubit control using single-flux
  quantum pulses,'' \emph{Phys. Rev. Applied}, vol.~6, no.~2, Aug. 2016.

\bibitem{supernpu}
K.~Ishida, I.~Byun, I.~Nagaoka, K.~Fukumitsu \emph{et~al.}, ``Super{NPU}: An
  extremely fast neural processing unit using superconducting logic devices,''
  in \emph{Proc. 53rd Annu. IEEE/ACM Int. Symp. Microarchit. (MICRO)}, Oct.
  2020, pp. 58--72.

\bibitem{nisq_parallel}
J.~Heckey, S.~Patil, A.~JavadiAbhari, A.~Holmes \emph{et~al.}, ``Compiler
  management of communication and parallelism for quantum computation,'' in
  \emph{Proc. 20th Int. Conf. Archit. Support Program. Lang. Oper. Syst.
  (ASPLOS)}, Mar. 2015, pp. 445--456.

\bibitem{coldflux}
C.~J. {Fourie}, K.~{Jackman}, M.~M. {Botha}, S.~{Razmkhah} \emph{et~al.},
  ``{ColdFlux} superconducting {EDA} and {TCAD} tools project: Overview and
  progress,'' \emph{{IEEE} Trans. Appl. Supercond.}, vol.~29, no.~5, Jan. 2019.

\bibitem{pasandi2019dynamic}
G.~Pasandi and M.~Pedram, ``A dynamic programming-based path balancing
  technology mapping algorithm targeting area minimization,'' in \emph{Proc.
  IEEE/ACM Int. Conf. Comput. Aided Des. (ICCAD)}, Nov. 2019.

\bibitem{pasandi2019efficient}
G.~Pasandi and M.~Pedram, ``An efficient pipelined architecture for
  superconducting single flux quantum logic circuits utilizing dual clocks,''
  \emph{{IEEE} Trans. Appl. Supercond.}, vol.~30, no.~2, Nov. 2019.

\bibitem{shahsavani2017integrated}
S.~N. Shahsavani, T.-R. Lin, A.~Shafaei, C.~J. Fourie, and M.~Pedram, ``An
  integrated row-based cell placement and interconnect synthesis tool for large
  {SFQ} logic circuits,'' \emph{{IEEE} Trans. Appl. Supercond.}, vol.~27,
  no.~4, 2017.

\bibitem{wrspice}
\BIBentryALTinterwordspacing
{Whiteley Research Inc.}, ``{WR}spice circuit simulator,'' 2017. [Online].
  Available: \url{http://www.wrcad.com/wrspice.html}
\BIBentrySTDinterwordspacing

\bibitem{jsim}
E.~S. Fang and T.~{Van Duzer}, ``A {Josephson} integrated circuit simulator
  ({JSIM}) for superconductive electronics application,'' in \emph{Extended
  Abstracts of 1989 Int. Supercond. Electronics Conf. (ISEC)}, Jun. 1989, pp.
  407--410.

\bibitem{pisq}
K.~Bertels, A.~Sarkar, and I.~Ashraf, ``Quantum computing--from {NISQ} to
  {PISQ},'' \emph{{IEEE} Micro}, vol.~41, pp. 24--32, Sep. 2021.

\bibitem{Barenco1995}
A.~Barenco, C.~H. Bennett, R.~Cleve, D.~P. {DiVincenzo} \emph{et~al.},
  ``Elementary gates for quantum computation,'' \emph{Phys. Rev. A}, vol.~52,
  no.~5, pp. 3457--3467, Nov. 1995.

\bibitem{transmon}
J.~Koch, T.~M. Yu, J.~Gambetta, A.~A. Houck \emph{et~al.}, ``Charge-insensitive
  qubit design derived from the {Cooper} pair box,'' \emph{Phys. Rev. A},
  vol.~76, no.~4, Oct. 2007.

\bibitem{quantum_eng}
P.~Krantz, M.~Kjaergaard, F.~Yan, T.~P. Orlando \emph{et~al.}, ``A quantum
  engineer's guide to superconducting qubits,'' \emph{Applied Physics Reviews},
  vol.~6, no. 2, 021318, Jun. 2019.

\bibitem{hutchings2017}
M.~D. Hutchings, J.~B. Hertzberg, Y.~Liu, N.~T. Bronn \emph{et~al.}, ``Tunable
  superconducting qubits with flux-independent coherence,'' \emph{Phys. Rev.
  Applied}, vol.~8, no.~4, Oct. 2017.

\bibitem{trappedidentical}
K.~R. Brown, J.~Kim, and C.~Monroe, ``Co-designing a scalable quantum computer
  with trapped atomic ions,'' \emph{npj Quantum Info.}, vol.~2, Nov. 2016.

\bibitem{qarch_hornibrook}
J.~M. Hornibrook, J.~I. Colless, I.~D. Conway~Lamb, S.~J. Pauka \emph{et~al.},
  ``Cryogenic control architecture for large-scale quantum computing,''
  \emph{Phys. Rev. Applied}, vol.~3, no.~2, Feb. 2015.

\bibitem{herr2011ultra}
Q.~P. Herr, A.~Y. Herr, O.~T. Oberg, and A.~G. Ioannidis, ``Ultra-low-power
  superconductor logic,'' \emph{Journal of applied physics}, vol. 109, no.~10,
  2011.

\bibitem{takeuchi2013adiabatic}
N.~Takeuchi, D.~Ozawa, Y.~Yamanashi, and N.~Yoshikawa, ``An adiabatic quantum
  flux parametron as an ultra-low-power logic device,'' \emph{Supercond. Sci.
  Technol.}, vol.~26, no.~3, 2013.

\bibitem{volkmann2013experimental}
M.~H. Volkmann, A.~Sahu, C.~J. Fourie, and O.~A. Mukhanov, ``Experimental
  investigation of energy-efficient digital circuits based on {eSFQ} logic,''
  \emph{{IEEE} Trans. Appl. Supercond.}, vol.~23, no.~3, Jan. 2013.

\bibitem{mukhanov2011energy}
O.~A. Mukhanov, ``Energy-efficient single flux quantum technology,''
  \emph{{IEEE} Trans. Appl. Supercond.}, vol.~21, no.~3, pp. 760--769, Jun.
  2011.

\bibitem{mcdermott2014}
R.~{McDermott} and M.~G. Vavilov, ``Accurate qubit control with single flux
  quantum pulses,'' \emph{Phys. Rev. Applied}, vol.~2, no.~1, Jul. 2014.

\bibitem{qarch_swamit2}
S.~S. Tannu, Z.~A. Myers, P.~J. Nair, D.~M. Carmean, and M.~K. Qureshi,
  ``Taming the instruction bandwidth of quantum computers via hardware-managed
  error correction,'' in \emph{Proc. 50th Annu. IEEE/ACM Int. Symp.
  Microarchit. (MICRO)}, Oct. 2017, pp. 679--691.

\bibitem{alphazero}
M.~Dalgaard, F.~Motzoi, J.~J. S{\o}rensen, and J.~Sherson, ``Global
  optimization of quantum dynamics with {AlphaZero} deep exploration,''
  \emph{npj Quantum Info.}, vol.~6, Jan. 2020.

\bibitem{nisq+}
A.~Holmes, M.~R. Jokar, G.~Pasandi, Y.~Ding \emph{et~al.}, ``{NISQ}+: Boosting
  quantum computing power by approximating quantum error correction,'' in
  \emph{Proc. ACM/IEEE 47th Annu. Int. Symp. Comput. Archit. (ISCA '20)}, May
  2020, pp. 556--569.

\bibitem{kitaev1997}
A.~Y. Kitaev, ``Quantum computations: algorithms and error correction,''
  \emph{Russian Mathematical Surveys}, vol.~52, no.~6, pp. 1191--1249, Dec.
  1997.

\bibitem{solovaykitaev}
C.~M. Dawson and M.~A. Nielsen, ``The {Solovay}-{Kitaev} algorithm,''
  \emph{Quantum Info. Comput.}, vol.~6, no.~1, pp. 81--95, Jan. 2006.

\bibitem{virtualz}
D.~C. {McKay}, C.~J. Wood, S.~Sheldon, J.~M. Chow, and J.~M. Gambetta,
  ``Efficient {Z} gates for quantum computing,'' \emph{Phys. Rev. A}, vol.~96,
  no.~2, Aug. 2017.

\bibitem{sfqdc}
V.~Kaplunenko, M.~Khabipov, V.~Koshelets, K.~Likharev \emph{et~al.},
  ``Experimental study of the {RSFQ} logic elements,'' \emph{{IEEE} Trans.
  Magn.}, vol.~25, no.~2, pp. 861--864, Mar. 1989.

\bibitem{jokarqce}
M.~R. Jokar, R.~Rines, and F.~T. Chong, ``Practical implications of {SFQ}-based
  two-qubit gates,'' in \emph{Proc. IEEE Int. Conf. Quantum Comput. Eng.
  (QCE)}, 2021, pp. 402--412.

\bibitem{pranav}
P.~Gokhale, A.~Javadi-Abhari, N.~Earnest, Y.~Shi, and F.~T. Chong, ``Optimized
  quantum compilation for near-term algorithms with {OpenPulse},'' in
  \emph{Proc. 53rd Annu. IEEE/ACM Int. Symp. Microarchit. (MICRO)}, Oct. 2020,
  pp. 186--200.

\bibitem{qarch_swamit3}
S.~S. Tannu and M.~K. Qureshi, ``Not all qubits are created equal: a case for
  variability-aware policies for {NISQ}-era quantum computers,'' in \emph{Proc.
  24th Int. Conf. Archit. Support Program. Lang. Oper. Syst. (ASPLOS)}, Apr.
  2019, pp. 987--999.

\bibitem{avgfidelity}
M.~A. Nielsen, ``A simple formula for the average gate fidelity of a quantum
  dynamical operation,'' \emph{Phys. Lett. A}, vol. 303, no.~4, pp. 249--252,
  Oct. 2002.

\bibitem{nufidelity}
J.~{Ghosh}, ``A note on the measures of process fidelity for non-unitary
  quantum operations,'' \emph{arXiv preprint}, Nov. 2011,
  \href{http://arxiv.org/abs/1111.2478}{arXiv:1111.2478}.

\bibitem{sgrape}
N.~Leung, M.~Abdelhafez, J.~Koch, and D.~Schuster, ``Speedup for quantum
  optimal control from automatic differentiation based on graphics processing
  units,'' \emph{Phys. Rev. A}, vol.~95, no.~4, Apr. 2017.

\bibitem{echo2}
S.~Sheldon, E.~Magesan, J.~M. Chow, and J.~M. Gambetta, ``Procedure for
  systematically tuning up cross-talk in the cross-resonance gate,''
  \emph{Phys. Rev. A}, vol.~93, no.~6, Jun. 2016.

\bibitem{echo1}
A.~D. C\'orcoles, J.~M. Gambetta, J.~M. Chow, J.~A. Smolin \emph{et~al.},
  ``Process verification of two-qubit quantum gates by randomized
  benchmarking,'' \emph{Phys. Rev. A}, vol.~87, no.~3, Mar. 2013.

\bibitem{pasandi2019balanced}
G.~Pasandi and M.~Pedram, ``Balanced factorization and rewriting algorithms for
  synthesizing single flux quantum logic circuits,'' in \emph{Proc. Great Lakes
  Symp. VLSI}, May 2019, pp. 183--188.

\bibitem{pasandi2018sfqmap}
G.~Pasandi, A.~Shafaei, and M.~Pedram, ``{SFQmap}: A technology mapping tool
  for single flux quantum logic circuits,'' in \emph{Proc. IEEE Int. Symp.
  Circuits Syst. (ISCAS)}, May 2018.

\bibitem{pasandi2019pbmap}
G.~Pasandi and M.~Pedram, ``{PBMap}: a path balancing technology mapping
  algorithm for single flux quantum logic circuits,'' \emph{{IEEE} Trans. Appl.
  Supercond.}, vol.~29, no.~4, Nov. 2019.

\bibitem{leiserson1991retiming}
C.~E. Leiserson and J.~B. Saxe, ``Retiming synchronous circuitry,''
  \emph{Algorithmica}, vol.~6, no.~1, pp. 5--35, 1991.

\bibitem{sfq5ee}
S.~K. Tolpygo, V.~Bolkhovsky, T.~J. Weir, A.~Wynn \emph{et~al.}, ``Advanced
  fabrication processes for superconducting very large-scale integrated
  circuits,'' \emph{{IEEE} Trans. Appl. Supercond.}, vol.~26, no.~3, 2016.

\bibitem{haolin}
H.~Cong, M.~Li, and M.~Pedram, ``An 8-bit multiplier using single-stage full
  adder cell in single flux quantum circuit technology,'' \emph{{IEEE} Trans.
  Appl. Supercond.}, vol.~31, no.~6, Jun. 2021.

\bibitem{sodor}
\BIBentryALTinterwordspacing
{UC Berkeley Architecture Research}, ``The sodor processor collection.''
  [Online]. Available: \url{https://github.com/ucb-bar/riscv-sodor}
\BIBentrySTDinterwordspacing

\bibitem{sfqpll}
D.~K. Brock and M.~S. Pambianchi, ``A 50 {GHz} monolithic {RSFQ} digital phase
  locked loop,'' in \emph{Proc. IEEE MTT-S Int. Microw. Symp. Digest (Cat. No.
  00CH37017)}, vol.~1, Jun. 2000, pp. 353--356.

\bibitem{qiskit}
M.~S. {ANIS}, H.~Abraham, {AduOffei}, R.~Agarwal \emph{et~al.}, ``Qiskit: An
  open-source framework for quantum computing,'' 2021.

\bibitem{murali2020}
P.~Murali, D.~C. {McKay}, M.~Martonosi, and A.~Javadi-Abhari, ``Software
  mitigation of crosstalk on noisy intermediate-scale quantum computers,'' in
  \emph{Proc. 25th Int. Conf. Archit. Support Program. Lang. Oper. Syst.
  (ASPLOS)}, Mar. 2020, pp. 1001--1016.

\bibitem{lloyd2018}
S.~Lloyd and C.~Weedbrook, ``Quantum generative adversarial learning,''
  \emph{Phys. Rev. Lett.}, vol. 121, no.~4, Jul. 2018.

\bibitem{Barends2016}
R.~Barends, A.~Shabani, L.~Lamata, J.~Kelly \emph{et~al.}, ``Digitized
  adiabatic quantum computing with a superconducting circuit,'' \emph{Nature},
  vol. 534, no. 7606, pp. 222--226, 2016.

\bibitem{Bernstein1997}
E.~Bernstein and U.~Vazirani, ``Quantum complexity theory,'' \emph{SIAM J.
  Comput.}, vol.~26, no.~5, pp. 1411--1473, Oct. 1997.

\bibitem{Cuccaro2004}
S.~A. Cuccaro, T.~G. Draper, S.~A. Kutin, and D.~P. Moulton, ``A new quantum
  ripple-carry addition circuit,'' \emph{arXiv preprint}, Oct. 2004,
  \href{http://arxiv.org/abs/quant-ph/0410184}{arXiv:quant-ph/0410184}.

\bibitem{Draper2006}
T.~G. Draper, S.~A. Kutin, E.~M. Rains, and K.~M. Svore, ``A logarithmic-depth
  quantum carry-lookahead adder,'' \emph{Quantum Info. Comput.}, vol.~6, no.
  4\&5, pp. 351--369, Jul. 2006.

\bibitem{JavadiAbhari}
A.~Javadi-Abhari, ``Towards a scalable software stack for resource estimation
  and optimization in general-purpose quantum computers,'' Ph.D. dissertation,
  Princeton University, 2017.

\bibitem{Prakash2019}
P.~Murali, A.~Javadi-Abhari, F.~T. Chong, and M.~Martonosi, ``Formal
  constraint-based compilation for noisy intermediate-scale quantum systems,''
  \emph{Microprocessors and Microsystems}, vol.~66, pp. 102--112, Mar. 2019.

\bibitem{grover}
L.~K. Grover, ``A fast quantum mechanical algorithm for database search,'' in
  \emph{Proc. 28th Annu. ACM Symp. Theory of Computing (STOC '96)}, Jul. 1996,
  pp. 212--219.

\bibitem{hertzberg2020}
J.~B. Hertzberg, E.~J. Zhang, S.~Rosenblatt, E.~Magesan \emph{et~al.},
  ``Laser-annealing {Josephson} junctions for yielding scaled-up
  superconducting quantum processors,'' \emph{npj Quantum Info.}, vol.~7, Aug.
  2021.

\bibitem{prakash}
P.~Murali, J.~M. Baker, A.~Javadi-Abhari, F.~T. Chong, and M.~Martonosi,
  ``Noise-adaptive compiler mappings for noisy intermediate-scale quantum
  computers,'' in \emph{Proc. 24th Int. Conf. Archit. Support Program. Lang.
  Oper. Syst. (ASPLOS)}, Apr. 2019, pp. 1015--1029.

\end{thebibliography}

\end{document}